\documentclass[12pt]{article}

\usepackage{amsmath,amssymb,color}
\usepackage{graphicx}

\input epsf.sty

\newcommand{\Bx}{x_{\rm B}}

\newcommand{\cQ}{ {\cal Q} }
\newcommand{\GeV}{{\rm GeV}}
\newcommand{\veps}{\mbox{\boldmath $\varepsilon$}}

\begin{document}

\begin{titlepage}

\centerline{\large\bf Generic modelling of non-perturbative quantities}
\vspace{2mm}
\centerline{\large\bf and a description of hard exclusive $\pi^+$ electroproduction}

\vspace{15mm}

\centerline{\bf C. Bechler and D. M\"uller}

\vspace{10mm}

\vspace{5mm} \centerline{\it Institut f\"ur Theoretische Physik
II, Ruhr-Universit\"at Bochum} \centerline{\it D-44780 Bochum,
Germany}

\vspace{1cm}

\centerline{\bf Abstract}

\vspace{0.5cm}
\noindent

Based on Regge-inspired arguments and counting rules, we formulate
empirical models for zero-skewness generalized parton
distributions (GPDs) ${\widetilde H}^{(3)}$ and ${\widetilde
E}^{(3)}$ in the iso-vector sector. If a hypothetical $a_1/\rho_2$
master trajectory is taken into account, we find that  the
polarized deep inelastic structure function $g_1^{(3)}$,
axial-vector form factor, pseudoscalar  form factor, and lattice
data are well described. Thereby, we use a symmetric valence
scenario in which the `spin puzzle' in the iso-vector sector is
trivially resolved. Utilizing a minimalist `holographic' GPD
principle, tying the $t$-channel angular momentum and collinear
conformal spin together, we build skewness dependent GPD models.
Confronting these models with  HERMES and JLAB measurements of
hard exclusive $\pi^+$ electroproduction within the collinear
factorization approach, we might conclude that minimalist GPD
models are disfavored at leading order. We provide then a set of
GPDs on the cross-over line that fairly describes both HERMES and
JLAB measurements.

\vspace{4cm}


\vspace{0.5cm}


\end{titlepage}

\tableofcontents

\newpage

\section{Introduction}
\label{Sec-Int}

\begin{figure}[t]
\begin{center}
\includegraphics[width=16cm]{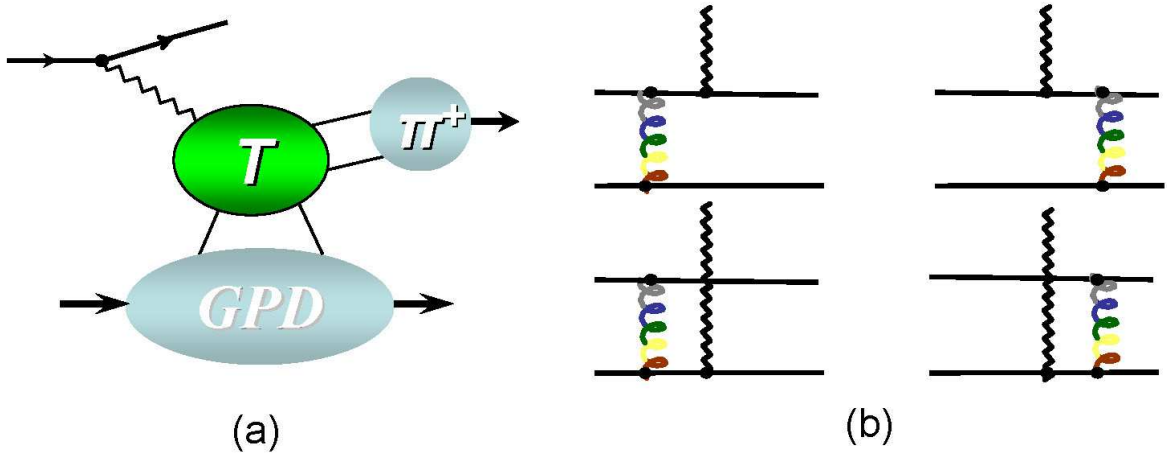}
\end{center}
\caption{\label{Factorization}
\small (a) Factorization of hard exclusive $\pi^+$ electroproduction at
large photon virtuality ${\cal Q}^2$ into a hard-scattering amplitude $T$,
the  flavor changing generalized parton distribution,
and  pion distribution amplitude $\phi_\pi$. (b) Hard scattering amplitude
at leading order.}
\end{figure}

The hard exclusive meson electroproduction off proton, e.g., for $\pi^+$
\begin{equation}
 e^-(k)\, p(P_1) \to e^-(k')\, n(P_2)\, \pi^+(q_2)\,,
\end{equation}
is considered as a promising channel to study the partonic content
of the nucleon. If the proton is probed with a {\em longitudinal}
polarized photon with sufficiently large virtuality ${\cal Q}^2 =
- (k-k^\prime)^2$,  the amplitude  factorizes in a hard part,
calculable in perturbation theory, and non-perturbative quantities
that are universal, i.e., process independent, however,
conventionally defined \cite{ColFraStr96}. As illustrated in
Fig.~\ref{Factorization} (a), the virtual photon knocks out a
quark-antiquark pair from the nucleon. In the collinear
factorization framework the probability {\em amplitude} for this
partonic process are the generalized parton distributions (GPDs),
which depend on the longitudinal momentum fraction and a non-zero
momentum flow in the $t$-channel
\cite{MueRobGeyDitHor94,Rad96,Ji96a}. The formation of the meson
is described within a  distribution amplitude (DA)
\cite{BroLep80,EfrRad80}, i.e., the minimal Fock state component
of the meson wave function, where the transverse  distance of the
quark-antiquark pair is squeezed.

The GPD framework is viewed as a powerful tool that provides
insight in the partonic content of the nucleon and it ties
perturbative and non-perturbative methods, for comprehensive
reviews see Refs.~\cite{Die03a,BelRad05}.  The hard exclusive
processes, described to leading order (LO) accuracy, allow to
access the GPDs on the cross-over line \cite{Ter05}. In
particular, the hard exclusive photon electroproduction is widely
considered as a theoretical clean process, see, e.g.,
Ref.~\cite{BelMueKir01} and references therein. The hard exclusive
meson electroproduction, which might be utilized as a flavor
filter, was elaborated at LO accuracy for numerous channels
\cite{Rad96a,ManPilWei97a,ManPilWei97,ManPilRad98} and based on
GPD models, estimates have been given, see, e.g.,
Refs.~\cite{VanGuiGui99,FraPobPolStr99,FraPolStrVan99}.
Perturbative next-to-leading order corrections were evaluated in
Refs.~\cite{BelMue01a,IvaSzyKra04,BelFreMue99}, and power
suppressed contributions were studied in
Refs.~\cite{ManPil99,Kiv01,Kiv05,AniIvaPirSymWal09}. Model studies
show  that perturbative corrections are rather large, see also the
comprehensive studies in Ref.~\cite{DieKug07}. Moreover, the
factorization breaks down for power suppressed contributions and
so a systematic expansion does not exist \cite{ColFraStr96} and
various  helicity amplitudes with leading twist-three suffer from
so-called end-point singularities \cite{ManPil99}.

The primary goal of revealing GPDs from the hard exclusive meson
electroproducion is rather challenging for both experiment and
theory. First, the cross section parts, which arise from
longitudinal and transverse polarized photon exchange, must be
separated.  To apply the collinear factorization framework, the
photon virtuality must be sufficiently large. Unfortunately, the
onset of the perturbative framework is controversially discussed,
where some considerations suggest that the perturbative
description becomes reliable at a scale of $\cQ^2 \gtrsim 10\,
\GeV^2$.  Such a scale is larger than the virtuality that is
accessible in present fixed target experiments. We  like to
emphasize that the emission and absorbtion of a quark from the
target resembles the Feynman mechanism and intuitively  this might
be considered as well describable. On the other hand the
formation of the meson can be affected to a larger extent by power
suppressed contributions, related to its transverse size. Relying
on this picture, improvements have been suggested that  provide
models rather than a systematic framework.

For the hard exclusive electroproduction of a longitudinal
polarized  $\rho^0_L$ meson  it has been demonstrated that in a
GPD  framework,  which contains also transverse degrees of
freedom, experimental measurements can be described
\cite{VanGuiGui99,GuiMor07,GolKro05,GolKro07}. In particular, at
small $\Bx$ the model \cite{GolKro05,GolKro07}, in which the GPD
part is treated in the collinear  approximation, successfully
predicts the spin density matrix.  However, from the success of
such models one cannot draw conclusions about the collinear GPD
approach, since model uncertainties are large and more importantly
the interplay between perturbative and non-perturbative
corrections in the collinear factorization approach and the
transverse degrees of freedom is essentially not understood. To
judge on the collinear factorization approach, one might confront
theoretical ``predictions'' with experimental measurements. The
problem of this approach is obvious, namely,  one relies rather on a
GPD model to reveal the GPD from experimental measurements.
Thereby, the non-perturbative GPD models, not calculable  from
first principles, suffer from ad hoc assumptions, see, e.g.,
discussion in Ref.~\cite{KumMuePas08}.

In this article we have a closer look to the  phenomenological
application of the GPD formalism  to the  hard exclusive $\pi^+$
electroproduction \cite{ManPilRad98,FraPobPolStr99,BelMue01a}.
More specifically, we consider the longitudinal photoproduction
\begin{eqnarray}
\gamma_{L}^\ast(q_1) p(P_1) \to
n(P_2) \pi^+(q_2)
\end{eqnarray}
within the collinear factorization approach at leading order of
perturbation theory. We like to confront this approach with cross
section measurements from the HERMES collaboration
\cite{Airetal07} and in HALL C at JLAB \cite{Horetal07,Bloetal08}.
In the later experiment the longitudinal cross section could be
separated, however, both the energy and the photon virtuality are
rather low. At the HERMES experiment a Rosenbluth separation cannot
be performed. Fortunately, according to Regge-inspired model
calculations \cite{GuiLagVan97,KasMurMos08,KasMos09}, the pion
pole enhances the longitudinal cross section at smaller values of
$-t$ and it becomes the dominate part. We also include in our
studies preliminary HERMES results for the  single transverse
proton spin asymmetry \cite{Hri08}.

Our goal is to confront GPD models with  measurements and to
illuminate the interplay of model dependence  and theoretical
uncertainties. To set up our zero-skewness GPD models, we follow
the idea, applied to polarized parton distribution functions
(PDFs) \cite{BroBurSch94}, to constrain or reveal non-perturbative
quantities from generic arguments. Our framework is set up in
Mellin space and it allows us to skew the resulting models in a
convenient way, where we utilize the $t$-channel SO(3) partial
wave (PW) expansion \cite{Pol98,JiLeb00,Die03a}.

The outline of the paper is as follows. In
Sect.~\ref{HadronicAmplitude} we recall the perturbative framework
for hard exclusive $\pi^+$ electroproduction at LO accuracy  in the
momentum fraction representation and then we present it  in
terms of conformal Mellin moments. In Sect.~\ref{Sec-Mod} we
explore generic GPD modelling with Mellin moments, where  we are
guided by Regge-inspired arguments and counting rules. Thereby, we
have to face the old problem that unnatural parity exchanges
challenge Regge phenomenology. We will inspect the meson spectrum
and we will argue that a odd $J^{++}$ leading Regge trajectory
might exist. As a side remark we shortly discuss  the so-called `spin
puzzle' and we employ a symmetric valence matching scheme of collective and
partonic degrees of freedom.
Our zero-skewness GPD models are then introduced and we compare them
with experimental measurements and lattice data. Relying on an ad
hoc assumption, we set up a minimalist GPD model and inspect its
properties.  In Sect.~\ref{Sec-ExpResGPD} we shortly recall the
illness of the perturbative framework and we utilize a scale setting
prescription that might cure it. We provide then  model
predictions  for the longitudinal cross section and single
transverse  proton  spin asymmetry and confront them with HERMES
and JLAB measurements.  We will then take the point of view that an
improved LO formalism is reliable and we present GPDs on the
cross-over line that describe experimental data. Finally, we
summarize and give our conclusions.

\section{Collinear factorization approach at LO}
\label{HadronicAmplitude}

We consider the hard exclusive  $\pi^+$  electroproduction off a
transverse polarized proton in  the target rest-frame. More
specifically, the $z$-axis is directed along the momentum of the
virtual photon and both the azimuthal angles  $\phi$ of the pion
momentum and $\phi_S$ of the proton polarization vector are
specified by the so-called Trento convention
\cite{BacDAlDieMil04}. The cross section is straightforwardly
evaluated in the collinear factorization approach to LO accuracy
\cite{ColFraStr96}. Thereby, the part  of the cross section that
arises from the exchange of a longitudinal polarized virtual
photon is counted as the leading twist-two contribution. The
exchange of a transverse polarized photon is on amplitude level
formally suppressed by $1/{\cal Q}$ and will not be considered.

In the following we can safely neglect the pion mass and mass
difference of proton and neutron, i.e., we set  $M \equiv M_p =
M_n$. In the twist expansion of the cross section we will adopt
here the same conventions as used for the hard exclusive photon
electroproduction in Ref.~\cite{BelMue08}. Namely, we decompose
first the amplitude by inserting photon helicity states into a
leptonic and hadronic part. We exactly calculate the leptonic part
and drop any power suppressed contribution in the hadronic part.
Thereby, we choose $1/{\cal Q}$ as expansion parameter and the
Bjorken variable $\Bx = \cQ^2/( 2 q_1 \cdot P_1)$ as scaling
variable. Finally, we can write the  differential cross section
for hard exclusive $\pi^+$ electroproduction, induced by
longitudinal polarized photon exchange, at twist-two level as
\begin{eqnarray}
\label{XsectionFinal}
\frac{d \sigma^{\pi^+}}{d {\cal Q}^2 d \Bx d t d \varphi}
\stackrel{{\rm Tw}-2}{=}
\frac{\alpha^2_{\rm em}}{{\cal Q}^8}
 \frac{\veps\, y^2 }{1- \veps}
\frac{\Bx}{\sqrt{1+\epsilon^2}}
\Bigg\{\!\!\!\!\!\!\!&&\!\!\!
(1-\xi^2) |\widetilde {\mathrm H}_{\pi^+}|^2
-\frac{t}{4 M^2} |\xi\widetilde {\mathrm E}_{\pi^+}|^2
- 2 \xi^2 \Re {\rm e}
\left( \widetilde {\mathrm H}_{\pi^+}
\widetilde {\mathrm E}_{\pi^+}^{\ast}\right)
\\
&&-
2  \sin(\phi-\phi_S)\,
\sqrt{\frac{t-t_{\rm min}}{t_{\rm min}}}\,  \xi^2\
\Im {\rm m}\left(
\widetilde {\mathrm H}_{\pi^+} \widetilde {\mathrm E}_{\pi^+}^{\ast}\right)
\Bigg\}\Bigg|_{\xi \to \frac{\Bx}{2-\Bx}}\,.
\nonumber
\end{eqnarray}
Here  $y = q_1 \cdot P_1/ k \cdot P_1$ is the photon energy
momentum fraction, $t=(P_2-P_1)^2$  is the squared momentum
transfer, the symmetric scaling variable $\xi$ is consequently
replaced by $\Bx/(2 - \Bx)$, and $\epsilon = \Bx M/\cQ$ is a
shorthand. The polarization parameter
\begin{equation}
\label{Def-PolPar}
\veps=
\frac{1-y-\frac{1}{4}y^2\epsilon^2
}{
1-y+\frac{1}{2}y^2+\frac{1}{4}y^2\epsilon^2}
\end{equation}
is the ratio of longitudinal and transverse photon flux.
The minimal value of $-t$ is
\begin{eqnarray}
-t_{\rm min} =
\cQ^2 \frac{2 (1-\Bx)(1-\sqrt{1+\epsilon^2})+\epsilon^2}{4\Bx(1-\Bx)+\epsilon^2}
\approx  \frac{M^2 \Bx^2}{1 - \Bx}\,,
\end{eqnarray}
where the indicated approximation is valid for $\Bx M^2/{\cal Q}^2
\ll 1-\Bx$. The hadronic part is contained in the {\em transition
form factors} (TFFs) $\widetilde {\mathrm H}_{\pi^+}$ and
$\widetilde {\mathrm E}_{\pi^+}$. They might be viewed as physical
quantities that depend on the symmetric scaling variable $\xi$,
$t$, and  $\cQ^2$.

As depicted in Fig.~\ref{Factorization}a, the TFF $\widetilde
{\mathrm H}_{\pi^+}$  ($\widetilde {\mathrm E}_{\pi^+}$)
factorizes into a convolution of a  hard amplitude $T_{u d}$ with
the GPD $\widetilde H^{u d}$ ($\widetilde E^{u d}$) and the pion
DA $\phi_\pi$\cite{ColFraStr96}:
\begin{eqnarray}
\label{MesonProFunctions}
\left\{\!\!
\begin{array}{c}
\widetilde {\mathrm H}_{\pi^+}
\\
\widetilde {\mathrm E}_{\pi^+}
\end{array}
\!\!\!\!
\right\}
= \int_0^1 du \int_{-1}^{1} d x \
\phi_\pi(u,\mu^2)
T_{ud}\left(
u, x, \xi|\alpha_s(\mu^2_r),\cQ^2/\mu^2,\cQ^2/\mu_r^2
\right)
\left\{\!\!
\begin{array}{c}
\widetilde H^{ud}
\\
\widetilde E^{ud}
\end{array}
\!\!
\right\}\!
\left( x, \xi, t|\mu^2 \right)
+ \dots\,,
\end{eqnarray}
where the ellipsis stands for power suppressed contributions.
Here, we distinguish between the renormalization scale $\mu_r$ and
the factorization scale $\mu$. Note that the truncation of the
perturbation series induces residual scale dependence. The
$\widetilde H^{u d}$ and $\widetilde E^{u d}$ GPDs, depending on
the quark momentum fraction $x$, $\xi$, $t$, and $\mu$, are flavor
non-diagonal and parity-odd.  They are defined by a form factor
decomposition of light-ray operator matrix elements in the
standard notation, see, e.g., Ref.~\cite{BelMue01a}. The leading
twist-two pion DA $\phi_\pi$, depending on the quark momentum fraction
$u$ and factorization scale $\mu$, is conventionally normalized to
the pion decay constant:
\begin{eqnarray}
\label{DA-Nor}
\int_0^1\!du\, \phi_\pi(u,\mu^2) =f_\pi\,, \quad f_\pi \simeq 133\,
\mbox{MeV}\,.
\end{eqnarray}

The short-distance dynamics, i.e., the hard amplitude $T_{u d}$,
can be perturbatively calculated as a series in the strong
coupling $\alpha_s(\mu_r)$, known to NLO accuracy \cite{BelMue01a}.
From the tree diagrams, presented in Fig.~\ref{Factorization}b,
we recover the well known result
\cite{ManPilRad98,FraPobPolStr99,VanGuiGui99}:
\begin{equation}
\label{LOamplitude}
T_{ud} (u, x, \xi|\alpha_s(\mu^2_r),\cQ^2/\mu^2,\cQ^2/\mu_r^2 )
= \frac{C_F}{N_c} \alpha_s(\mu^2_r)
\left\{
\frac{Q_u}{(1 - u) \left( \xi - x - i 0 \right)}
-
\frac{Q_d}{u \left( \xi + x - i 0 \right)}
\right\}
+
{\cal O} (\alpha_s^2)\,,
\end{equation}
where $C_F=4/3$, $N_c=3$, and the fractional quark charges are
$Q_u = 2/3$ and $Q_d = -1/3$. The hard amplitude (\ref{LOamplitude})
is proportional to the running coupling
\begin{eqnarray}
\label{Def-alpha}
\alpha_s(\mu^2_R) =
\frac{4\pi}{\beta_0 \ln\frac{\mu^2_R}{\Lambda^2_{\rm QCD}}}
\quad\mbox{with}\quad
\beta_0\equiv 11-2 n_f/3 = 9
\,,
\end{eqnarray}
where we set $n_f$, the number of active quarks, to three and in
the following we will take the QCD scaling parameter $\Lambda_{\rm
QCD}=200\,\,\text{MeV}$. We emphasize that in principle the
renormalization scale $\mu_R$ can be ambiguously chosen because
this ambiguity is annulled in the full perturbative series.

In the following we will stay at LO accuracy and equate the
factorization scale with the photon virtuality, i.e., we set
$\mu^2 = \cQ^2$.  In the LO approximation the hard amplitude
(\ref{LOamplitude}) possesses a factorized $u$ and $\xi\mp x$
momentum fraction dependence%
\footnote{Note that although NLO corrections
might be rather large, it is expected that this factorization is
only slightly spoiled by radiative NLO corrections,
see, e.g.,~Ref.~\cite{Mue98}.}. Since the
pion DA has definite charge  parity-even, possessing the
symmetry $\phi_\pi(u)= \phi_\pi(1-u)$, the convolution integral
\begin{eqnarray}
\label{Def-ConInt}
{\cal I}_\pi(\cQ^2) = \frac{1}{3 f_\pi}\int_{0}^{1}\! du \,
 \frac{\phi_{\pi}(u,\cQ^2)}{u},
 \end{eqnarray}
enters as overall factor in the normalization of the TFFs
(\ref{MesonProFunctions}). The value of the inverse moment
(\ref{Def-ConInt}) is directly `measurable' at LO accuracy in the
fusion process $\gamma^\ast \gamma\to \pi^0$ \cite{BroLep80}. The
factorized GPD part of the $\widetilde{\mathrm H}^{ud}$ and
$\widetilde{ \mathrm E}^{ud}$ TFFs \eqref{MesonProFunctions}
might be decomposed in terms with definite charge parity,
\begin{eqnarray}
\widetilde{H}^{ud\pm}(x,\cdots)=
\widetilde{H}^{ud}(x,\cdots) \pm \widetilde{H}^{ud}(-x,\cdots)\,,
\quad
\widetilde{E}^{ud\pm}(x,\cdots)=
\widetilde{E}^{ud}(x,\cdots) \pm \widetilde{E}^{ud}(-x,\cdots)\,.
\end{eqnarray}
Here the $\pm$ superscript refers in the first place to the
symmetry behavior under $x\to -x$ and it coincides for both of
our parity-odd GPDs with the $t$-channel charge parity $C= \pm 1$.
Employing isospin symmetry, one can express the flavor
off-diagonal GPDs to the common flavor diagonal ones in the
iso-vector sector \cite{ManPilWei97}:
\begin{equation}
\label{SU2symm}
\widetilde{H}^{ud} =
\widetilde{H}^{(3)}  \equiv \widetilde{H}^{u} - \widetilde{H}^{d} \,,
\quad
\widetilde{E}^{ud} =
 \widetilde{E}^{(3)} \equiv \widetilde{E}^{u} - \widetilde{E}^{d} \,.
\end{equation}
The convolution of these GPDs with the LO hard scattering kernel
\begin{eqnarray}
\label{Def-CFF}
{\cal F}(\xi,t,\cQ^2) \stackrel{\rm LO}{=}
\int_{-1}^{1}\! dx \, \frac{F(x,\xi,t,\cQ^2)}{\xi-x -i 0}\,,
\quad {\cal F}(F) = \left\{
\widetilde{\cal H}^{ud\pm} (\widetilde{H}^{ud\pm}),
\widetilde{\cal E}^{ud\pm}(\widetilde{E}^{ud\pm})
\right\}
\end{eqnarray}
might be denoted as  `Compton form factors' (`CFFs'). Note,
however, that only the charge even part (plus superscript) enters
in the LO approximation of the deeply virtual Compton scattering
(DVCS) amplitude. In the case of hard exclusive $\pi^+$ production
we have the combination
\begin{eqnarray}
\label{CCF-dec}
\left\{\!{
\widetilde{\cal H}^{ud}
\atop
\widetilde{\cal E}^{ud}
}\!\right\}
=
\left\{\!{
\widetilde{\cal H}^{(3)+}  + \frac{1}{6} \widetilde{\cal H}^{(3)-}
\atop
\widetilde{\cal E}^{(3)+}  + \frac{1}{6} \widetilde{\cal E}^{(3)-}
}\!\right\}
\end{eqnarray}
in which also the charge even part dominates. Putting all
definitions together, we finally have a very compact LO expression
for the TFFs
\begin{eqnarray}
\label{FFH}
\left\{
{\widetilde{\mathrm H}_{\pi^+}
\atop
\widetilde{\mathrm E}_{\pi^+} }
\!\!\right\}(\xi, t,\cQ^2) \stackrel{\rm LO}{=}
\alpha_{s}(\mu_r^2) \frac{2  f_\pi}{3}   {\cal I}_\pi(\cQ^2)
\left\{{
\widetilde{\cal H}^{(3)+} +  \frac{1}{6} \widetilde{\cal H}^{(3)-}
\atop
\widetilde{\cal E}^{(3)+}  +  \frac{1}{6} \widetilde{\cal E}^{(3)-}
}\!\right\}\!(\xi, t,\cQ^2) + \cdots\,.
\end{eqnarray}
Here the renormalization scale is ambiguous and so
$\alpha_{s}(\mu_r^2)$ might be loosely viewed as an ``effective
charge''.  As we will argue below in Sect.~\ref{Sec-ExpResGPD},
its value might be extracted from the measurement of the pion form
factor.  Let us emphasize that in principle Eq.~(\ref{FFH}) can
serve as a {\em qualitative} test of the collinear factorization
approach, where the TFFs of hard exclusive $\pi^+$
electroproduction are expressed in terms of quantities that can be
addressed in the LO approximation of the pion-to-photon transition
form factor, pion form factor,  and to a certain extent in DVCS.

Let us also introduce the conformal PW expansion of the TFFs. The
underlying symmetry is the collinear conformal group
SL(2,${\mathbb R}$) and in pQCD it can be used to diagonalize the
evolution equation, see, e.g., Ref.~\cite{BraKorMue03}.  The
conformal PW expansion for the pion DA is well known at LO
\cite{BroLep80,EfrRad80},
\begin{equation}
\label{PionDAexpansion}
\phi_{\pi}(v,\cQ^2)=
f_\pi\sum_{n=0 \atop {\rm even}}^\infty 6 (1-v)v C_n^{3/2}(2v-1)\,
{\mathbb E}_n(\mathcal{Q},\mathcal{Q}_0)
a_n(\cQ^2_0)\,,
\end{equation}
where the conformal PWs  are given in terms of orthogonal
Gegenbauer polynomials $C_n^{\nu}$ of order $n$ and index
$\nu=3/2$. The partonic ``quantum number'', conjugated to the
momentum fraction $u$ is the (integral) conformal spin $n+2$. The
first coefficient $a_0=1$  is fixed by the normalization
(\ref{DA-Nor}) and does not evolve, i.e., ${\mathbb
E}_0(\mathcal{Q},\mathcal{Q}_0)=1$. If the scale $\cQ > \cQ_0$
increases, the evolution operator
\begin{equation}
{\mathbb E}_n(\mathcal{Q},\mathcal{Q}_0) =
\left(
\frac{
\ln(\mathcal{Q}^2/ \Lambda_{\rm QCD}^2)
}{
\ln(\mathcal{Q}_0^2/\Lambda_{\rm QCD}^2)
}
\right)^{-\gamma_n^{(0)}/\beta_0}\,,
\quad
\gamma_n^{(0)}=
\frac{4}{3} \left(4 S_{n+1}-\frac{2}{(n+1)(n+2)}-3 \right)\,,
\end{equation}
reduces the magnitude of conformal moments $a_n$ with $n>0$. Here
the harmonic sum is defined as $S_m= \sum_{k=1}^{m}\frac{1}{k}$.
The inverse moment, given in Eq.~(\ref{Def-ConInt}) as convolution
integral, might be now expressed as a series of conformal moments
\begin{equation}
\label{PionDAexpansion-1}
{\cal I}_\pi(\cQ^2)=
\sum_{n=0 \atop {\rm even}}^\infty
{\mathbb E}_n(\mathcal{Q},\mathcal{Q}_0)  a_n(\cQ^2_0)\,.
\end{equation}

For the $t$-channel crossed GPDs, i.e., the so-called generalized
DAs, one might write down an analogous conformal PW expansion as
in Eq.~(\ref{PionDAexpansion}). The conformal moments of a quark
GPD are group theoretically defined in terms of Gegenbauer
polynomials $ C_j^{3/2}$ of order $j$ and with index $\nu=3/2$:
\begin{eqnarray}
\label{Def-ConGPDMom}
\left\{{\widetilde{H}_j \atop \widetilde{E}_j }\right\}(\eta,t,\cQ^2)
=
\frac{\Gamma(3/2) \Gamma(1+j)}{2^{j+1} \Gamma(3/2+j)}
\int_{-1}^1\! dx\,
\eta^j C_j^{3/2}(x/\eta)
\left\{{ \widetilde{H}
\atop
\widetilde{E}}\right\}(x,\eta,t,\cQ^2)\,.
\end{eqnarray}
Note that our definition ensures that [anti]symmetric functions
possesses  only even [odd] integral moments, which are even
polynomials in $\eta$ of order $j$ [or $j-1$]. In the limit
$\eta\to 0$ the conformal moments of $\widetilde H$ GPD coincide
with the  Mellin moments of a polarized PDF:
\begin{eqnarray}
\label{Def-ComMelMom}
\widetilde{H}_j^{\pm}(\eta=0,t,\cQ^2) =
\frac{1 \pm (-1)^j}{2} \int_{0}^1\! dx\,
x^j \widetilde{H}_j^{\pm}(x,\eta=0,t,\cQ^2)\,.
\end{eqnarray}
As in the case of a  SO(3) PW expansion, labelled by the
$t$-channel angular momentum, such a series does not converge for
the $s$-channel analog. We can borrow us the known recipes  and
employ a Sommerfeld-Watson transform \cite{MueSch05}, which reads
for the `CFFs' as follows
\begin{eqnarray}
\label{FFMellinmoments-+}
\left\{ {\widetilde{\cal H} \atop \widetilde{\cal E}}\right\}^{\pm}
\!\!\! \stackrel{\rm LO}{=}
\frac{1}{2 i} \int_{c-i \infty}^{c +i \infty} dj \,\,
\frac{\Gamma(5/2+j)}{\Gamma(3/2)\Gamma(3+j)} (\xi/2)^{-j-1}
{\mathbb E}_j(\cQ,\cQ_0)
\frac{\mp 1- e^{- i \pi j}}{\sin{\pi j}}
\left\{ {\widetilde{H}_j
\atop
\widetilde{E}_j}\right\}^{\pm}\!\!\!\!(\xi,t,\cQ_0^2)\,.
\end{eqnarray}
The analytic continuation of the evolution operator is
straightforward by replacing $n\to j$ and representing the
harmonic sum $S_{j+1} = \psi(j+2)-\psi(1)$ in terms of Euler's
$\psi$ functions. The analytic continuation of integral conformal
GPD Mellin moments (\ref{Def-ConGPDMom}), is dictated by the
requirement that after crossing $\xi \to 1/\xi$, the Mellin-Barnes
integral yields a series, rather similar to the convolution
integral (\ref{PionDAexpansion-1}). In general this requires a
separate continuation of even and odd conformal moments.

\section{Generic modelling in  Mellin space}
\label{Sec-Mod}

A GPD $F(x,\eta,t)$ at a given input scale $\mu^2$ is an intricate
function of three variables, the $s$-channel momentum fraction
$x$, its $t$-channel counterpart $\eta \sim (P_1 - P_2)_+$, and
$t$.  With the growing amount of experimental data for the hard
exclusive photon electroproduction it became obvious that popular
ad hoc GPD models do not posses predictive power on quantitative
level; a more detailed discussion can be found in
Ref.~\cite{KumMuePas08}. For our purpose, namely, a qualitative
confrontation of the collinear factorization approach with hard
exclusive $\pi^+$ electroproduction measurements, it is sufficient
to deal with zero-skewness GPD models and an ad hoc prescription
to skew them. In the following we work with conformal GPD Mellin
moments (\ref{Def-ConGPDMom}), which in principle allows us to set
up flexible GPD models.

To parameterize the $\eta$-dependence of GPDs, we follow
Ref.~\cite{Pol98} and expand the conformal moments
(\ref{Def-ConGPDMom})  in terms of $t$-channel SO(3) PWs. These
PWs are given in terms of the Wigner rotation matrices
$d^J_{0,0}(\cos\theta)$ and $d^J_{0,1}(\cos\theta)$, i.e., by
Legendre or Gegenbauer polynomials, respectively
\cite{Die03a,KumMuePas07}. Here the second subscript refers to the
helicity difference of the final state $N^{\uparrow (\downarrow )}
\bar{N}^{\uparrow}$ in the $t$-channel center-of-mass frame, i.e.,
for (anti-)aligned nucleon helicities it is  zero (one). The
center-of-mass scattering angle $\theta$ is approximatively
related to the skewness parameter $\eta \approx -1/\cos\theta$.
Note that the exact relation would yield more cumbersome formulae
that  provide no advantages or improvements for our GPD modelling.
The crossed version of the Wigner rotation matrices are noted for
shortness as $\hat d^J_{0,0}(\eta)$ and $\hat d^J_{0,1}(\eta)$ and
they are normalized in the limit $\eta \to 0$ to one
\cite{KumMuePas07}. Their analytic continuation with respect to
the $t$-channel angular momentum $J$ is defined in the traditional
way and we express them by hypergeometric functions:
\begin{eqnarray}
\label{PW01}
\hat d^J_{0,1}(\eta) &\!\!\!=\!\!\!&
\frac{\Gamma(3/2)\Gamma(2+J)}{2^{J} \Gamma(1/2+J)}\, \eta^{J-1}\,
{_2F_1}\!\left(\!{-J+1, J+2 \atop 2} \Big| \frac{\eta-1}{2\eta}\!\right)\,,
\\
\label{PW00}
\hat d^J_{0,0}(\eta) &\!\!\!=\!\!\!&
\frac{\Gamma(1/2)\Gamma(1+J)}{2^{J} \Gamma(1/2+J)}\, \eta^{J}\,
{_2F_1}\!\left(\!{-J, J+1 \atop 1} \Big| \frac{\eta-1}{2\eta}\!\right)\,.
\end{eqnarray}

Now we can straightforwardly write down the SO(3)-PW expansion for
integral conformal GPD moments (\ref{Def-ConGPDMom}). For the
target helicity conserved GPD $\widetilde H$ the SO(3)-PW
expansion reads
\begin{eqnarray}
\label{PWE-tH}
\widetilde H^\pm_j(\eta, t) =
\sum_{J = 1}^{j+1}  \frac{1\mp (-1)^J}{2}
\widetilde H^\pm_{jJ-1}(t)\,  \eta^{j+1-J}\, \hat d^J_{0,1}(\eta)\,.
\end{eqnarray}
Here even $j$ moments (plus superscript) and  odd $j$ moments
(minus superscript) are build up by odd $J^{++}$ and
even $J^{--}$ exchanges, respectively \cite{JiLeb00,Die03a}:
\begin{eqnarray}
\label{tra-J++odd&tra-J--even}
J^{\rm PC}=1^{++},\,3^{++},\,\cdots,\, (j+1)^{++}\,,
\quad
J^{\rm PC}=2^{--},\,4^{--},\,\cdots,\, (j+1)^{--}\,.
\end{eqnarray}
Note that here the difference $j-J$ is always odd and so a
confusion with the notion signature, which might be used in both
the conformal and SO(3) PW expansion, could appear. We
consequently stick here and in the following to the Regge terminology
and associate signature plus and minus with even and odd $J$,
respectively.

The $t$-channel helicity conserved GPD moments are expressed by a
combination of $s$-channel flip and non-flip contributions:
\begin{eqnarray}
\label{PWE-tHE}
\left[
\widetilde H^\pm_j+\frac{t}{4 M^2} \widetilde E^\pm_j
\right]\!(\eta, t)
=
\sum_{J = 0}^{j}  \frac{1\pm (-1)^J}{2}
\left[
\widetilde H^{\prime^\pm}_{jJ}(t) +\frac{t}{4 M^2}
\widetilde E^\pm_{jJ}(t)
\right]  \eta^{j-J}\, \hat{d}^J_{0,0}(\eta)\,.
\end{eqnarray}
Now the  even $j$ moments (plus superscript) and odd $j$ moments
(minus superscript) are associated with even  $J^{-+}$ and odd
$J^{+-}$ $t$-channel exchanges, respectively \cite{JiLeb00,Die03a}:
\begin{eqnarray}
\label{tra-J-+even&tra-J+-odd}
J^{\rm
PC}=0^{-+},\,2^{-+},\,\cdots,\, j^{-+}\,,
\quad
J^{\rm PC}=1^{+-},\,3^{+-},\,\cdots,\, j^{+-}\,.
\end{eqnarray}
We might utilize the crossed version of Wigner's rotation
matrices $\hat{d}^J_{0,0}(\eta)$ within the assignment
(\ref{tra-J-+even&tra-J+-odd}) in the SO(3) PW expansion of
$\widetilde E$ GPD conformal moments:
\begin{eqnarray}
\label{PWE-E}
\widetilde E^\pm_j(\eta, t) =
\sum_{J = 0}^{j}  \frac{1\pm (-1)^J}{2}
\widetilde E^\pm_{jJ}(t)\, \eta^{j-J}\, \hat{d}^J_{0,0}(\eta)\,.
\end{eqnarray}
Then the $\widetilde H$ GPD conformal moments can be alternatively
represented by the SO(3) PW expansion (\ref{PWE-tH}) and
(\ref{PWE-tHE}) in terms of the amplitudes $\widetilde
H_{j,J-1}(t)$ and $\widetilde H^\prime_{jJ}(t)$, respectively.
Consequently, these both PW amplitudes can be mapped to each
other. Below we will assume that the ${\widetilde H}_j$ conformal
moments are only built by one leading SO(3) PW  $J=j+1$ with odd
$J^{++}$ or even $J^{--}$. This implies that all PW amplitudes of
$\widetilde H^{\prime}_{jJ}(t)$, appearing in Eq.~(\ref{PWE-tHE})
and containing contributions from even  $J^{-+}$ or odd $J^{+-}$
exchanges, are exited. Of course, they can be entirely expressed
in terms of $\widetilde H_{j,j}(t)$ amplitudes:
\begin{eqnarray}
\label{Con-HHprime}
\widetilde H_{jJ}^\prime(t) =
\frac{\Gamma(J+3/2)\Gamma(j+1)}{2^{J-j} \Gamma(j+3/2)\Gamma(J+1)}
\widetilde H_{jj}(t)\,,\;\;   j\ge J\ge 0\,.
\end{eqnarray}

In the following  sections we  model the zero-skewness Mellin
moments, $$\widetilde{H}^{(3)}_j(\eta=0,t) =
\widetilde{H}^{(3)}_{jj}(t)\quad\mbox{and}\quad
\widetilde{E}^{(3)}_j(\eta=0,t) = \widetilde{E}^{(3)}_{jj}(t),$$
in the iso-vector sector. Although Regge theory is developed for
on-shell scattering, it is obvious from the phenomenology that
{\em effective} Regge behavior also appears in off-shell
amplitudes. As a fact of matter, in the collinear factorization
approach {\em effective} Regge behavior is encoded in  PDFs,
obtained from global fits, and so also in GPD models. Moreover, it
is rather robust under evolution, except for the ``pomeron''-like
exchange \cite{CudDonLan99}. Hence, Regge poles should be a
building block for our partonic SO(3) PW amplitudes. The
corresponding PW amplitudes will be interpreted according to Regge
phenomenology and they are in the $t$-channel, i.e., $\gamma_L^*
\pi^- \rightarrow n \bar{p}$, given by exchanges with definite
$J$. Analogous as in Regge theory, applied to high-energy on-shell
scattering, this allows us to resum $t$-channel exchanges and,
finally, they determine the small-$x$ behavior of GPDs or PDFs.
Moreover, a link between {\em effective} Regge behavior and
$t$-dependence of nucleon form factors has been also employed or
observed in zero-skewness GPD models
\cite{DieFelJakKro04,GuiPolRadVan04,MueSch05}. We are going along
the line of Ref.~\cite{MueSch05,KumMuePas07,KumMue09} and
unpublished work, shortly mentioned in Ref.~\cite{KumMuePas08a}
(see corresponding talk at DIS2008). There the $t$-dependence of
$H$ and $E$ GPD Mellin moments is modelled as a product of leading
and daughter Regge poles. Since the empirical $H$ and $E$ GPD
models describe within generic numbers, obtained from counting
rules and Regge theory, form factors, unpolarized PDFs, and
lattice data, we are encouraged to utilize generic modelling in
the following sections.

However, first we have a look to the phenomenological status of
Regge trajectories in the meson spectrum. We then build our
zero-skewness GPD models and confront them with the polarized deep
inelastic scattering (DIS) structure function $g_1$, form factor
measurements, and lattice data. Finally, we skew our GPD models
within a {\em minimalist} prescription as they arise from the
dominance of the leading SO(3) PW in the expansion (\ref{PWE-tH})
and (\ref{PWE-E}). We also discuss the  model properties, relevant
for the phenomenology of hard exclusive $\pi^+$ electroproduction.

\subsection{Regge trajectories from  meson spectroscopy}
\label{Sec-RegThe}

\begin{figure}[t!]
\centering
\includegraphics[width=16cm]{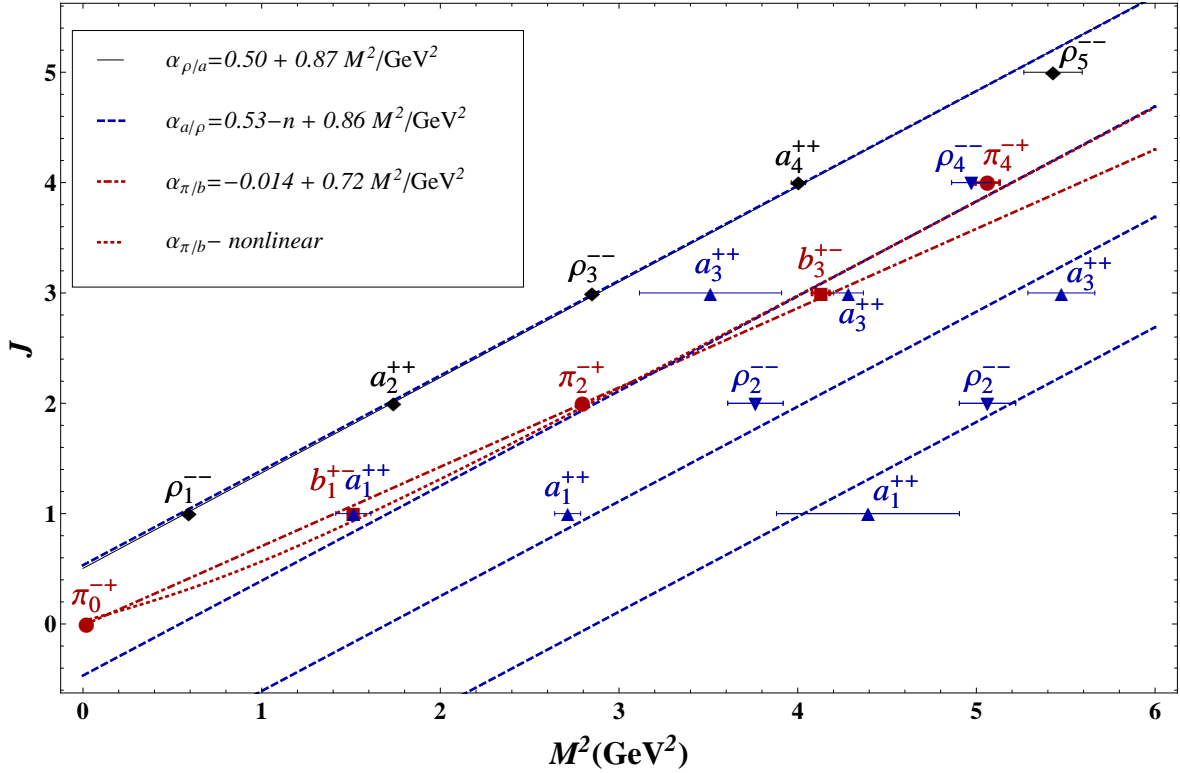}
\caption{\label{figRegTra} \small
Spectrum of iso-vector mesons
$a_1$, $a_3$, $\rho_2$ and $\rho_4$, the states of the leading
$\pi_0/b_1$ and $\rho_1/a_2$ trajectories. Curves show linear
Regge trajectory fits, $a_1/\rho_2$ (dashed), $\pi_0/b_1$
(dot-dashed) and $\rho_1/a2$ (solid), and an example of a
non-linear  $\pi_0/b_1$ trajectory (dotted). Spectroscopic data are
from the Review of Particle Physics~\cite{RPPs08}.}
\end{figure}

According to the quantum numbers (\ref{tra-J++odd&tra-J--even})
and (\ref{tra-J-+even&tra-J+-odd}), we are dealing with unnatural
parity $$P= -(-1)^J$$ $t$-channel exchanges such as $\pi^{-+}_0/b^{+-}_1$
and $a_1^{++}/\rho^{--}_2$ ones. Such exchanges challenge Regge
phenomenology, applied to high energy on-shell scattering
processes, see reviews~\cite{IrvWor77,Col77}. We are not aware
that the problems there, e.g., an unexpected large $a_1$ pole
contribution in $\pi^- p \to \rho^0 n$ \cite{IrvWor77}, were
finally solved, e.g., in terms of very large Regge cut
contributions. Therefore, and because of the immense experimental
progress in meson spectroscopy during the last decade, see the
comprehensive reviews~\cite{Bug04,KleZai07}, it is worth to have a
fresh look to Regge trajectories in the meson spectrum. Thereby, we take
experimental data from the Particle Data Group (PDG) \cite{RPPs08},
where we also include mesons listed under further states.

According to Eq.~(\ref{tra-J++odd&tra-J--even}), the leading SO(3)
PW amplitudes $\widetilde H^{(3)+}_{jj}$ and $\widetilde
H^{(3)-}_{jj}$  with even and odd $j=J-1$, respectively, carry odd
$J^{++}$ and even $J^{--}$ $t$-channel quantum numbers. We
associate them with meson exchanges that belong to $a_1$ and
$\rho_2$ trajectories. In the Chew-Frautschi plot
in Fig.~\ref{figRegTra} the iso-vector mesons
$$a_1,\;\; a_3,\;\; \rho_2\,\;\; \mbox{and}\;\;  \rho_4$$
might lie on trajectories, which  we traditionally
consider as linear equidistantly  spaced, i.e.,
\begin{eqnarray}
J(M^2)=  \alpha^{(n)}(M^2)\,,\quad
\alpha^{(n)}(M^2) \simeq \alpha_0-n + \alpha^\prime M^2\,,
\quad n= 0,1,2,\cdots,
\label{RegTra-lin}
\end{eqnarray}
where $\alpha(M^2)\equiv \alpha^{(n=0)}(M^2)$ is called
master trajectory. In particular, the states
$$
a_1(1260)[1230\pm 40],\;\;
a_3(2070)[2070\pm 30],\;\;
\rho_4(2240)[2230\pm 25]
$$
[values in square brackets are the mass and its error in MeV,
quoted by the PDG] might belong to the trajectory with the largest
intercept, called $a_1$ trajectory.  Note however, that $\rho_2$
is not established and
$$a_3(1875)[1874\pm 43 \pm 96]$$
lies between the $\rho_1/a_2$ master  and the $a_1$
trajectory within an equal distance of slightly more than
1-$\sigma$ standard deviation (errors are added in quadrature).
The states $a_1(1640)$, $\rho_2(1940)$, $a_3(2310)$ and
$a_1(2095)$, $\rho_2(2240)$ might belong to two daughter
trajectories.

Let us first suppose that both states $a_3(1875)$ and $a_3(2070)$
nearly belong to the $a_1$ trajectory.  A least square fit within
a linear $a_1$ trajectory and  two daughters reveals that
$\chi^2/{\rm d.o.f.}\approx 4.1/(9-2)$ does not contradict our
hypothesis and the result
\begin{eqnarray}
\label{Reg-a1}
\alpha^{\rm est}_{a_1}(s)
=
- (0.32 \pm  0.30) + \left(0.83 \pm 0.07\right)\, s/{\rm GeV}^2\,,\qquad
\chi^2/{\rm d.o.f.}\approx 0.59
\end{eqnarray}
confirms the  $a_1$ trajectory that is {\em ``established''} in
the literature. The trajectory (\ref{Reg-a1}) has a typical
$t$-slope and its intercept $ -0.6 \lesssim \alpha_{a_1}(0)
\lesssim 0$ suffers from a large uncertainty. Such an uncertainty
appears also in the analyzes of high-energy scattering processes,
see, e.g., discussion in Ref.~\cite{EllKar88}.

Certainly, other interpretations of the $a_J$ meson spectrum are
not excluded, e.g., one might consider the $a_1$ trajectory as
degenerated with a non-linear $\pi_0$ trajectory \cite{Afo07}, as
illustrated  by the dotted curve in Fig.~\ref{figRegTra}. Also it
has been argued that chiral symmetry might be partially restored
at large  $J$, and so {\em exotic} parity doubling%
\footnote{ Using  $J= \alpha^{(n)}(M^2)$ and the assignment
$P=(-1)^{L+1}$ and $C=(-1)^{L+S}$, where according to the quark
model $S=\{0,1\}$, it is easy to see that a parity doubling {\em
pattern} appears in the meson spectrum, except for the master
trajectory $\alpha_{\rho_1/a_2}(M^2)$ itself. Partial restoration
of chiral symmetry predicts that parity $(-1)^{J}$ states on the
${\rho_1/a_2}$ master trajectory at sufficiently large $J$ are
degenerated with two further  parity $(-1)^{J+1}$ states, carrying
$C=+1$ and $C=-1$, respectively.} appears in the meson spectrum,
see review \cite{Afo07b}. According to that, at sufficiently large
$J$ a meson with natural parity, belonging to the master
trajectory, should posses two partners with unnatural parity.  The author
of Ref.~\cite{Swa03} argued that the $\rho^{--}_3(1690)$ meson
has a $3^{++}(\sim 1700)$ partner. The large mass error allows us to
speculate that  $a_3^{++}(1875)$, reported in 2002 in Ref.~\cite{Chu02},
is a parity doubling candidate, which  might indicate
a partial chiral symmetry restoration.

We  now  count the  $a_3^{++}(1875)$ state to a master trajectory
and like to know whether it is degenerated with the $\rho_1/a_2$
trajectory. A fit to the $a_1/\rho_2$ meson spectrum yields the result
\begin{eqnarray}
\label{Reg-a1-dau}
\alpha^{(n)}_{a_1/\rho_2}(s) =
(0.53 \pm  0.34) -n + \left(0.86 \pm 0.08\right)\, s/{\rm GeV}^2\,,
\qquad
\chi^2/{\rm d.o.f.}\approx 0.67\,,
\end{eqnarray}
which is with $n=1$ fully compatible with the previous one
(\ref{Reg-a1}). Although the uncertainty is large, the means
somehow suggest that this trajectory is degenerated with the
$\rho_1/a_2$ one.  In Fig.~\ref{figRegTra}. our fit result
(\ref{Reg-a1-dau}), shown as dashed curve, is hardly to
distinguish from the $\rho_1/a_2$ master trajectory (solid curve),
which we obtained from a separate fit to its first five members.

Further support for the existence of a master $a_1/\rho_2$
trajectory arises from polarized DIS. Here, the rise of the
iso-vector polarized nucleon structure function $g^{(3)}_1(\Bx)$
at small-$\Bx$ might be viewed as a $\Bx^{-\lambda}$ power
behavior with $\lambda\sim 1/2$ \cite{Bas05,KuhCheLea09},
illustrated below in the left panel of Fig.~\ref{figPolDis}. Such
a power behavior, partly even a stronger one,  is included in
standard PDF parameterizations and it is predicted by a double log
resummation \cite{BarErmRys95}. On the other hand in the Regge
framework, one equates $\lambda = \alpha_{a_1}(0)$ \cite{Hei73}.
Hence, from the {\em ``established''} $a_1$ trajectory
(\ref{Reg-a1}) it is then expected that $g^{(3)}_1(\Bx)$ in the
small-$\Bx$ region vanishes or approaches a constant, which
contradicts experimental measurements. As the authors of
Refs.~\cite{Bas05,KuhCheLea09}, we are not aware that this  Regge
puzzle is solved.

Although it is not conclusive from the meson spectrum that a
leading trajectory for unnatural parity states exists,
we conjecture a  master  $a_1/\rho_2$-trajectory,
\begin{eqnarray}
\label{Reg-?}
\alpha_{a_1/\rho_2}(s)  \cong
\alpha_{\rho_1/a_2}(s)=
(0.5\pm 0.05) + (0.87\pm 0.05)\, s/{\rm GeV}^2\,,
\qquad \chi^2/{\rm d.o.f.} = 0.24\,,
\end{eqnarray}
to explain the rise of the polarized DIS structure function
$g_1^{(3)}$. We consider this hypothetic Regge trajectory  as
degenerated with the master $\rho_1/a_2$-trajectory and we view
the {\em ``established''} $a_1$-trajectory as its first daughter.
We did not investigate whether our conjecture resolves problems or
contradicts phenomena in on-shell processes at high energy
\cite{IrvWor77,Col77}.

We inspect now the even $J^{-+}$ and odd $J^{+-}$ Regge
trajectories that appear in the SO(3) PWs of ${\widetilde
E}^+_{jj}$ and ${\widetilde E}^-_{jj}$, respectively, where $j=J$,
see Eq.~(\ref{tra-J-+even&tra-J+-odd}). Presumably, the leading
trajectories are  non-linear and their daughters are populated
with natural parity vector states $\pi_1^{-+}$, having exotic
quark model quantum numbers ($S=0$ and even $L$). Since this is
not compatible with our  hypothesis (\ref{RegTra-lin}), we do not
utilize daughter trajectories.  The leading signature even and
odd trajectory are populated with
$$
\pi_0^\pm\, [(139.57018 \pm 0.00035)\,{\rm MeV}]\,,\quad
\pi_2(1670)\, [(1672.4 \pm 3.2)\,{\rm MeV}]\,,\quad
\pi_4(2250)\, [(2250 \pm  15)\,{\rm MeV}]
$$
and
$$b_1(1235)\, [(1229.5\pm 3.2)\,{\rm MeV}]\,,\quad
b_3(2025)\, [(2032\pm 12)\, {\rm MeV}]
$$
mesons, respectively.

If we ignore the $\pi_4(2250)$ state, reported in 2001 in
Ref.~\cite{Anietal01}, and we assume that the trajectories are
linear and degenerated, we find from a fit the  $\pi_0/b_1$-trajectory
\begin{eqnarray}
\label{Def-RegTraPi}
\alpha_{\pi_0/b_1}(s) = (-0.014 \pm 0.002) + (0.72\pm 0.01 )\,
s/{\rm GeV}^2\,, \qquad  \chi^2/{\rm d.o.f.}  \approx 1.1\,,
\end{eqnarray}
shown in Fig.~\ref{figRegTra} as  dot-dashed line. Since this
trajectory is often used, we denote it as {\em ``established''}.
Compared to a typical meson Regge trajectory,  the {\em
``established''} $\pi_0/b_1$-trajectory has a smaller slope
parameter and much smaller intercept, which is close to zero.
Remarkably, the intercept can be numerically expressed as
$\alpha_\pi(0) = - m_\pi^2\alpha^\prime_\pi$. This allows us to
parameterize the pion trajectory in terms of the pion mass and the
slope parameter:
\begin{eqnarray}
\label{Reg-Tra-pi}
\alpha_{\pi_0}(s) = -\alpha^\prime_\pi (m_\pi^2- s)\,
\;\;
\mbox{with} \;\;
m_\pi= \sqrt{\frac{|\alpha_\pi|}{\alpha^\prime_\pi}} \approx 0.14\,\GeV\,.
\end{eqnarray}

However, the {\em ``established''} trajectory (\ref{Def-RegTraPi}) is
disfavored by the fifth state $\pi_4(2250)$ (a fit within a linear
trajectory to all five states provides $\chi^2/{\rm d.o.f.}
\approx 3.5$). From Fig.~\ref{figRegTra} it seems to be obvious
that the anomalous behavior of the trajectory appears at low $J$
values rather than large ones, see for example curve (dotted) and compare
also with master trajectory. If we now adapt the heavy pion
world ($m^2_\pi \sim 600\,{\rm MeV}^2$), the $\pi_0/b_1$
trajectory  (\ref{Reg-Tra-pi}) could be viewed as linear and
parallel to the master one. On the other hand in the chiral limit
$m_\pi\to 0$ the pion  might be viewed as a singular state that does
not belong to a linear $b_1/\pi_2$-trajectory, connecting the
other four states.

Presumably, the non-linearity of the $\pi_0/b_1$-trajectory is
related to spontaneous chiral symmetry breaking. However, we
believe that the chiral limit does not provide a perfect guidance
to model builders that are interested to describe the real world.
Thus, we should in fact utilize a non-linear pion trajectory
rather than a pion pole. However, for simplicity  we compromise and
choose the hypothetic $a_1/\rho_2$- and $\pi_0/b_1$-trajectory to
be linear within a common slope parameter that is smaller and larger
than those of the {\em ``established''} $a_1/\rho_2$- and
$\pi_0/b_1$-trajectory, respectively,
\begin{eqnarray}
\label{Reg-Par-pi}
\alpha(t)= \alpha+ \alpha^\prime t\,,\;\;
\alpha =\frac{1}{2}\,,\;\alpha^\prime =\frac{4}{5\,\GeV^2}\,;
\quad
\alpha_\pi(t)= \alpha_\pi+ \alpha^\prime t \,,\;\;
\alpha_\pi = - m_\pi^2 \alpha^\prime\,,\; m_\pi= 0.14\,\GeV\,.
\end{eqnarray}
Here and in the following we denote a leading master trajectory as
$\alpha(t)$ and its intercept with the shorthand $$\alpha \equiv
\alpha(t=0).$$

\subsection{Modelling  $ \widetilde H_{jj}^{(3)}$}
\label{subsec-ModH}

\subsubsection{$t=0$ case: polarized PDFs}

We consider first the $t=0$ case of the leading SO(3) PW
amplitudes in conformal GPD moments. They  coincide then with the
common Mellin moments of the polarized PDF:
\begin{eqnarray}
\label{Def-PW-H-fl}
\widetilde{H}_{jj}^{(3)\pm}(t=0) \equiv \Delta q^{(3)\pm}_j
\quad\mbox{for}\quad
j= \left\{
{\mbox{even}\, (+) \atop \mbox{odd}\,\,\, (-) }
\right. \,,
\end{eqnarray}
cf.~Eq.~(\ref{Def-ComMelMom}).
For even and odd $j$ the parity charge-even and -odd type PDFs are to be taken
\begin{equation}
\label{FjDelta0}
\Delta q^{(3)\pm}_j =\int_0^1 dx \,\, x^j \Delta q^{(3)\pm}(x)\,,
\quad \Delta q^{(3)\pm} = \Delta u  - \Delta d \pm
\left(\Delta \bar{u}  - \Delta \bar{d} \right),
\quad j = \left\{ {\mbox{even}\, (+) \atop \mbox{odd}\,\,\, (-) } \right..
\end{equation}
We also distinguish between valence, sea, and anti-quarks, where the
sea content of polarized quarks is expressed by the anti-quarks
$\Delta\bar{q}(x)$. The polarized charge parity-even PDF is then
given as sum of valence quarks and antiquarks and the charge parity-odd one
is equated to valence quarks%
\footnote{
For convenience, we proceed in momentum fraction space and below in
Sect.~\ref{subsubsec-t-dec-Reg} we return to Mellin space.
}:
\begin{eqnarray}
\label{Dec-val/sea}
\Delta q^+(x) = \Delta q^{\rm val}(|x|) + \Delta q^{\rm sea}(|x|)\,,
\quad
\Delta q^{\rm sea}(|x|)= 2 \Delta \bar{q}(|x|)\,,
\quad
\Delta q^{-}(x)= {\rm sign}(x)\,\Delta q^{\rm val}(|x|)\,.
\end{eqnarray}

Let us make a side remark with respect to the `spin puzzle', which
we understand as the mismatch of quark model and partonic degrees
of freedom. In the quark model the nucleon
possesses no antiquarks  and its constituents are considered as collective
degrees of freedom. Loosely spoken, the constituent quarks are
somehow made up  from the partonic content. If one literally
identifies the partonic and quark model degrees of freedom, one
easily generates momentum fraction or spin `puzzles'. It might be
somehow legitimate to identify the collective constituent quarks
with the partonic valence quarks \cite{GluRey77} and one might
wonder whether the `spin puzzle' exists then. The `spin puzzle' is
perhaps easily solved by an appropriate matching prescription of
collective and partonic degrees of freedom or, as in the case of
the `momentum fraction puzzle', one simply accept that one deals
with different degrees of freedom. Of course, the decomposition
of the proton spin in terms of its fundamental degrees of freedom
remains  a serious challenge.

Although it is rather likely
that SU(6) symmetry in polarized valence quark PDFs holds true at
small-$x$ values and it is broken in the large-$x$ region, we find
it interesting to explore the valence
matching scheme of collective and partonic  degrees of freedom.
To do so, we adapt flavor-spin SU(6) [SU(3) $\otimes$ SU(2)]
symmetry to fix the normalization of polarized valence quarks in
the iso-vector sector by the group theoretical value of the
axial-vector coupling $g_A^{\rm val}$:
\begin{eqnarray}
\label{NorDelqm}
g_A^{\rm val}\equiv \int_0^1 \Delta q^{(3)-}(x)
&\!\!\! = \!\!\!&
u^{\uparrow,\rm val} - u^{\downarrow,\rm val}
- d^{\uparrow,\rm val}+d^{\downarrow,\rm val}
\nonumber\\
&\!\!\!  \stackrel{\rm SU(6)}{=} \!\!\!&
 \frac{4}{3} - \frac{-1}{3} = \frac{5}{3}.
\end{eqnarray}
The axial-vector charge $g_A$, i.e., the expectation value of the
axial-vector current in the forward kinematics, is given by the
lowest moment of the polarized charge parity-even quark PDFs.
Hence, both valence and sea quarks contribute:
\begin{eqnarray}
\label{NorDelq3}
g_A\equiv \int_0^1 \Delta q^{(3)+}(x) &\!\!\! = \!\!\!&
u^{\uparrow,\rm val} - u^{\downarrow,\rm val}
- d^{\uparrow,\rm val}+d^{\downarrow,\rm val}
+ 2 \left(\overline{u}^{\uparrow} - \overline{u}^{\downarrow}
-  \overline{d}^{\uparrow} + \overline{d}^{\downarrow}\right)
\nonumber\\
&\!\!\! \stackrel{\rm exp}{=} \!\!\!& 1.2670\pm 0.0035 \,,
\end{eqnarray}
and the experimental value, taken from Ref.~\cite{RPPs08}, is
smaller than the  SU(6) quark model expectation. For an isospin
symmetric polarized sea, the axial-vector charge $g_A$ reduces to
its valence part $g_A^{\rm val}$.  In such an ad hoc scenario,
historically used for simplicity, e.g., in standard PDF fits,
the SU(6) value (\ref{NorDelqm}) contradicts the experimental
finding (\ref{NorDelq3}). We
clearly spell out that the SU(6) value (\ref{NorDelqm}) is not
ruled out and a `spin puzzle' does not exist. Rather SU(6) symmetry
forces us to introduce an isospin asymmetric
sea with a large negative net contribution
\begin{eqnarray}
g_A^{\rm sea}\equiv \int_0^1\!dx\; \Delta q^{(3) {\rm sea}}(x)
 &\!\!\! = \!\!\!& 2\left(
\overline{u}^{\uparrow} -\overline{u}^{\downarrow}
-  \overline{d}^{\uparrow} +\overline{d}^{\downarrow}
\right)
\nonumber\\
&\!\!\! \stackrel{\rm SU(6)}{=} \!\!\!&  -\frac{2}{5}\,.
\end{eqnarray}
Here we used that $g_A=5/3-2/5 \approx 1.267$ is the mean of the
experimental value (\ref{NorDelq3}). As we will see below, this
SU(6) symmetric valence scenario is also not ruled out by
polarized DIS measurements and some lattice simulations. Our
scenario is in conflict with chiral quark soliton model
($\chi$QSM) considerations
\cite{DiaPetPobPolWei96,WakKub98,DreGoePolSchStrWei99,Goeetal00,Wak03},
predicting a flavor asymmetric sea within a positive value for
$g_A^{\rm sea}$; however, qualitatively supported by meson cloud
model considerations, suggesting a (small) negative value of
$g_A^{\rm sea}$ \cite{FriSch98,BorKai99,KumMiy01,CaoSig01}.

The next ingredient for our generic PDF model are large-$x$
counting rules, obtained from dimensional counting or diagrammatic
analysis \cite{BlaBro74,Gun73}. They state that for a proton with
helicity $+1/2$ a quark PDF vanishes for $x\to1$ as
$(1-x)^\gamma$, where the power
\begin{eqnarray}
\gamma=2 n_s -1+ (1-2h)
\end{eqnarray}
is determined by the number of spectators $n_s$ and the helicity
$h =\pm 1/2$ of the struck quark \cite{BroBurSch94}. For valence
and sea quarks with  helicity aligned to the proton one the
exponent $\gamma$ is given by $\beta = 3$ and $\beta = 7$,
respectively. If the struck quark helicity is anti-aligned these
values increase by $\delta\beta=2$ to $\beta+\delta\beta$. We
recall that the small-$x$ behavior of our polarized PDF is
determined by the intercept of the hypothetical master trajectory
(\ref{Reg-?}), i.e., $\lim_{x\to 0}\Delta q(x) \propto
x^{-\alpha}$, which we set to the generic value (\ref{Reg-Par-pi}),
$\alpha=1/2$ .

The small and  large-$x$ behavior are incorporated in a factorized
ansatz:
\begin{eqnarray}
\label{PDFmodel1}
\Delta q^{(3)} (x) &\!\!\!\equiv \!\!\!&
\widetilde H^{(3)} (x,\eta=0,t=0)
\\
&\!\!\!=\!\!\!&
g_A \, x^{-\alpha} \frac{(1-x)^{\beta}  -
\widetilde{h}\,(1-x)^{\beta+\delta \beta}
}{
B(1-\alpha,1+\beta)-  \widetilde{h}\, B(1-\alpha,1+\beta+\delta \beta)
}\,,
\nonumber
\end{eqnarray}
where the normalization of the lowest Mellin moment is given by
the corresponding partonic axial-vector couplings $g_A^{\rm val}$
and $g_A^{\rm sea}$. In the partonic model interpretation the
$\widetilde{h}$ parameter adjusts the ratio of quarks with aligned
and anti-aligned helicities. We might assume that  at small-$x$
the sea quark density becomes helicity independent, i.e., we set
$\widetilde{h}^{\rm sea}=1$. For this specific value the polarized
PDF (\ref{PDFmodel1}) behaves with $\alpha =1/2$  in the small-$x$
region as $x^{1-\alpha}=x^{1/2}$, which is nothing but the Regge
behavior that belongs to the established $a_1$ trajectory,
cf.~Eq.~(\ref{Reg-a1}), \cite{BasLan94}. However, the partonic
interpretation of $\widetilde{h}^{\rm sea}$  might be too rigid for
$\widetilde{h}^{\rm val}$. Namely, at large-$x$ also a certain
amount of valence quarks with aligned helicities might be
suppressed by some additional powers of $(1-x)^{\Delta\beta}$
\cite{BroBurSch94}. We choose $\Delta\beta=\delta\beta =2$ and
consider now $\widetilde{h}^{\rm val}$ as an effective parameter.
In the following considerations we set $\widetilde{h}^{\rm val}
\stackrel{\rm phe}{=}0.85$. Relying on our partonic inspiration,
the hypothetical Regge intercept,  SU(6) symmetric valence scenario,
and canonical counting rules, we can fix the model parameters:
\begin{eqnarray}
\label{Par-Set-H}
\alpha\!\stackrel{\rm gen}{=}
\frac{1}{2}\,,\;\;\alpha^\prime\stackrel{\rm gen}{=}\frac{4}{5}\,,
&&
g_A^{\rm val} \stackrel{\rm SU(6)}{=}
\frac{5}{3}\,,\phantom{-}
\quad
\widetilde{h}^{\rm val} \stackrel{\rm phe}{=} 0.85\,,
\quad
\beta^{\rm val} \stackrel{\rm gen}{=} 3\,,
\quad
\delta^{\rm val} \beta\stackrel{\rm gen}{=} 2\,,
\\
&&
g_A^{\rm sea} \stackrel{\rm SU(6)}{=} -\frac{2}{5}\,,
\quad
\widetilde{h}^{\rm sea} \stackrel{\rm gen}{=} 1\,,\phantom{.85}
\quad
\beta^{\rm sea} \stackrel{\rm gen}{=} 7\,,
\quad
\delta^{\rm sea} \beta\stackrel{\rm gen}{=} 2\,,
\nonumber
\end{eqnarray}
where  also the generic Regge slope parameter, see
Eq.~(\ref{Reg-Par-pi}), is included.

A few comments are in order. For our convenience we take the
parameters, specified in (\ref{Par-Set-H}), at the input scale
$\mu^2 = 2\,\GeV^2$.  The phenomenological values for $\beta$ and
$\delta\beta$ should in accordance with evolution be moderately
larger while the hypothetical Regge intercept $\alpha$ is rather
robust under evolution. Instead of a SU(6) symmetric valence
scenario we could also have adopted a standard one from the
literature. After an appropriate change of $\widetilde{h}^{\rm
val/sea}$, all of these scenarios allow us to describe the
measurements, discussed in this article. We add that the model can
be easily extended to unpolarized PDFs. In such a framework the
parameters $\widetilde{h}^{\rm val/sea}$, considered as effective
ones, can be constrained from the large-$x$ behavior of DIS
structure functions. It is beyond the scope of this paper to
present such studies.

The iso-vector part of the polarized structure function, expressed
at LO by the polarized PDF, can be experimentally obtained from
DIS measurements on proton and neutron target:
\begin{eqnarray}
g_1^{(3)}(\Bx,\cQ^2) &\!\!\!\equiv\!\!\!&
g_1^{(p)}(\Bx,\cQ^2)-g_1^{(n)}(\Bx,\cQ^2)
\\
&\!\!\!\stackrel{\rm LO}{=}\!\!\!&
\frac{1}{6} \Delta q^{(3)+}(x=\Bx,\mu^2=\cQ^2)\,.
\nonumber
\end{eqnarray}
Here we conventionally set the scale $\mu^2=\cQ^2$.
\begin{figure}[t!]
\centering
\includegraphics[width=16cm]{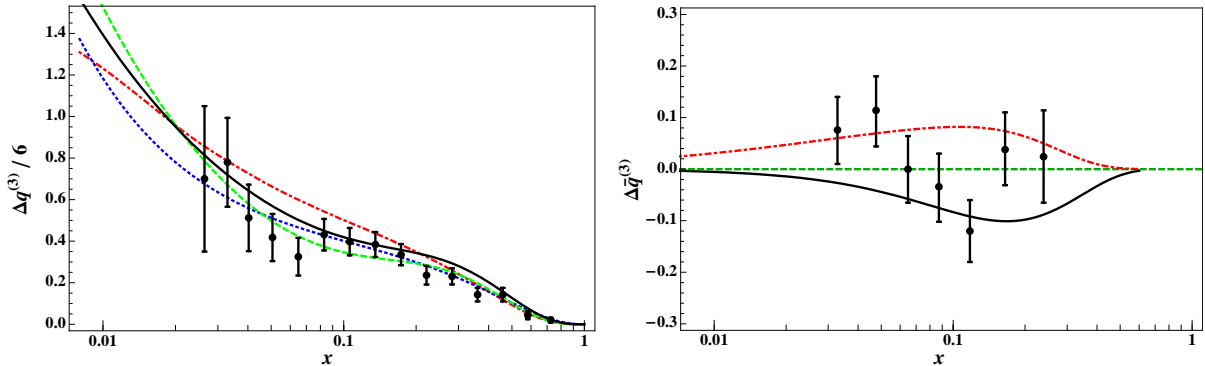}
\caption{\label{figPolDis}\small
Standard LO PDF parameterizations, described in the text, and a
generic PDF model (solid), specified in
Eqs.~(\ref{PDFmodel1}) and (\ref{Par-Set-H}), are confronted with
HERMES results \cite{Airetal07a, Airetal04} for the DIS structure
function $g^{(3)}_1(x)$  (left)  and  $(\Delta \overline{u}-
\Delta \overline{d})(x)$  (right), where only statistical errors
are shown.}
\end{figure}
Our generic PDF model fairly agrees with standard polarized PDF
parameterizations, taken at the input scale $\mu^2=2\, \GeV^2$.
This is illustrated in the left panel of Fig.~\ref{figPolDis},
where our model (solid) is displayed together with measurements of the
HERMES collaboration \cite{Airetal07a}, and standard
parameterizations: Gehrmann/Stirling A (95) \cite{GehSti95}
(dashed), Bl\"umlein/B\"ottcher (02) scenario 1 \cite{BluBot02}
(dotted), and Gl\"uck/Reya/Stratmann/Vogelsang (00) `broken
valence' scenario \cite{GluReyStrVog00} (dot-dashed). Note that we
fixed in all PDF parameterizations $\cQ^2 = 2\,\GeV$ and that the
various parameterizations have a rather different partonic
interpretation. The GS and BB parameterizations rely on a flavor
symmetric polarized sea, the `broken valence'  GRSV scenario
possesses a positive sign for the polarized sea in the iso-vector
sector, contrarily, to our SU(6) symmetric valence scenario which
requires a negative sign. Hence, the valence and sea quark content
of the axial-vector charge is rather different, too. Also the
small-$x$ behavior of the various parameterizations differ. Since
the areas under the PDF curves in the left panel of
Fig.~\ref{figPolDis} are fixed by the Bjorken sum rule (or
equivalently the normalization condition (\ref{NorDelq3}) for the
axial-vector current), the small-$x$ behavior of the various
parameterizations is essentially a consequence of the assumed
large-$x$ behavior and the specific parameterization, in
particular, the flavor scheme for the polarized sea.

Semi-inclusive pion production allows addressing polarized
antiquarks in the iso-vector sector. In the right panel of
Fig.~\ref{figPolDis} we show the partonic interpretation of a
HERMES measurement \cite{Airetal04}, obtained from a LO fit (full
circles), together with various PDF parameterizations. The
`measurement' suggests that $x(\Delta \overline{u}- \Delta
\overline{d})(x)$ is highly non-symmetric, where its sign might
even change in different $x$ regions; however, statistically, the
asymmetry is also compatibly with zero. The figure also
illustrates that present experimental data do not allow to
quantify the polarized sea and that even a qualitative judgement
on the different scenarios can not be drawn. This conclusion is
based on more solid ground, namely, on the extensive analysis of
polarized PDF uncertainties in global fits from
Ref.~\cite{FloSasStrVog09}. Our SU(6) symmetric valence scenario
is supported by global fits from Ref.~\cite{FloNarSas05}.

Because of the polarized PDF uncertainties, one might wonder
whether the so-called `spin puzzle' or 'crisis'  in the iso-vector
sector has ever existed. The variety in the behavior of  standard
polarized PDF parameterizations might be also rather important for
GPD model builders. Namely, these PDF uncertainties might propagate
into GPD model predictions for hard exclusive processes.

\subsubsection{$t$-decoration via Regge trajectories}
\label{subsubsec-t-dec-Reg}

To decorate the conformal GPD moments (\ref{Def-PW-H-fl}) with
$t$-dependence, we rely on dimensional counting  rules for the
large $-t$ behavior of form factors. Lattice results clearly show
that  with increasing $j$, i.e., in the momentum fraction
representation in the limit $x\to 1$, the $t$-dependence dies out
\cite{Hagetal03,Gocetal03}. To unify these features with Regge
behavior, we  write the $t$-dependence as product of monopoles,
build up from the leading Regge trajectory and its daughters:
\begin{equation}
\label{Hpartialwave-1}
\widetilde{H}_{jj}^{(3)}(t)=
\frac{\left(1+j-\alpha\right)_p}{\left(1+j-\alpha(t)\right)_p} \Delta q^{(3)}_j\,,
\quad
\alpha(t) \stackrel{\rm gen}{=} \frac{1}{2} + \frac{4\, t}{5\, \GeV^2}
\,,
\end{equation}
where the Pochhammer symbol for integral $p > 0$ is defined as
$$
(x)_p = x (x+1)\cdots (x+p-1)
$$
and the intercept  $\alpha$ has the same value as in the polarized
PDF (\ref{PDFmodel1}). Hence, the zeros of the pre-factor in our
model (\ref{Hpartialwave-1}) cancel the corresponding poles in the
PDF moments, which are then replaced by a product of $t$-dependent
Regge poles $1/[1+j-\alpha(t)]\cdots[p+j-\alpha(t)]$. The number
$p$ determines the asymptotic of the PW amplitudes
(\ref{Hpartialwave-1}) at $-t \to \infty$, which behave as
$(-t)^{-p}$. Employing the Drell-Yan exclusive-inclusive relation
$\gamma =2 p -1$ \cite{DreYan69}, we should choose $p=2$ and $p=4$
for valence ($\gamma=3$) and sea quarks ($\gamma=7$),
respectively.  To simplify the model procedure, we take for both
contributions $p=2$  and we complete the Regge intercept to a
generic leading meson trajectory (\ref{Par-Set-H}).

\begin{figure}[t!]
\centering
\includegraphics[width=18cm]{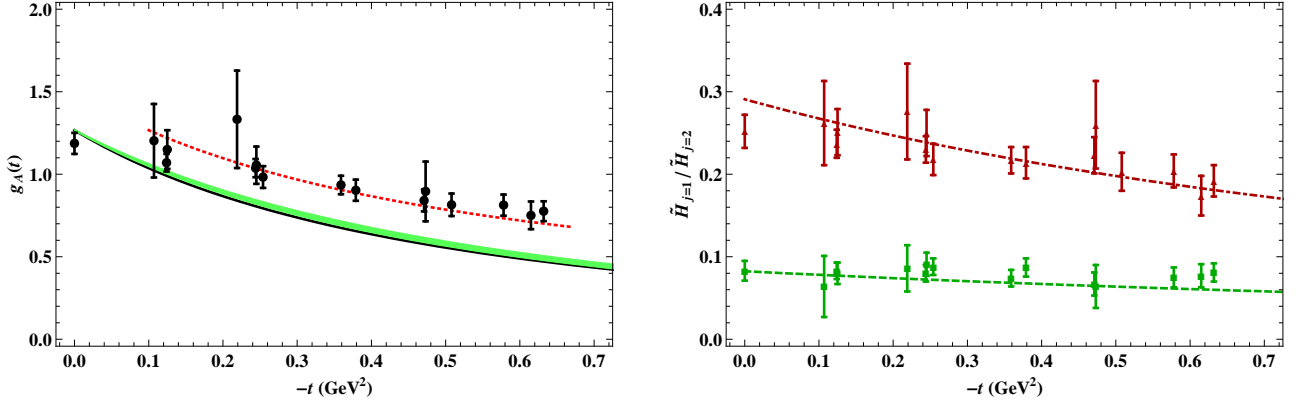}
\caption{\label{figparset}\small
The left panel shows the axial-vector form factor versus $-t$:
world fit (\ref{WorFit-gA}) [error band], generic $\widetilde H$
GPD model (\ref{Hpartialwave-1})  [solid], as specified in the
text, and lattice results (full circles).  In the right panel the
same GPD model is confronted with lattice data for generalized
form factors: $j=1$ (dot-dashed, triangles) and  $j=2$ (dashed,
squares). The scale is $\mu^2=4\, \GeV^2$ and lattice measurements
are taken from data set VI of Ref.~\cite{Hagetal07}.}
\end{figure}
We illustrate in the left panel of Fig.~\ref{figparset} that the
$t$-dependence of the monopole product ansatz
(\ref{Hpartialwave-1}) for $j=0$,
displayed as solid curve, is in fair agreement with world
dipole fits (error band),
\begin{eqnarray}
\label{WorFit-gA}
g_A(t)=\frac{g_A(0)}{\left(1-\frac{t}{M_A^2}\right)^2}, \qquad
M_A^2=(1.069\pm0.016) \,\GeV^2\; \left[(1.026\pm 0.021) \,\GeV^2\right],
\end{eqnarray}
to the axial-vector form factor measurements from $\pi^+$
electroproduction  and  neutrino scattering experiments [$M_A^2$
given in square brackets], taken from Ref.~\cite{Lieetal99}. We
also display lattice data (full circles) from the set VI of
Ref.~\cite{Hagetal07}, which at larger $-t$ considerables overshoot
the experimental measurements (only statistical errors are shown)
and possess at low $-t$ a rather flat $t$-dependence. The dotted
curve is included for illustration and it arises from a shift of
our solid model curve. In the right panel of Fig.~\ref{figparset}
we plot the second (dot-dashed) and third (dashed) moment
(\ref{Hpartialwave-1}) versus $-t$ for $\mu^2=4\, \GeV^2$, where
the polarized PDF moments are specified in Eq.~(\ref{Par-Set-H})
at $\mu^2=2\, \GeV^2$.  Our generic model predictions are now
compatible with lattice data. We emphasize that in particular
the lattice result for the $j=1$ valence generalized form factor
at $t=0$, given as%
\footnote{Lattice measurements suffer from systematical
uncertainties that can  hardly  be estimated and so it is common
to quote only the statistical error.} $0.25\pm 0.02$, does not
``rule out'' the SU(6) symmetric valence scenario prediction
$\approx 0.29$. In standard PDF parameterizations the $j=1$ Mellin
moment is typically smaller, namely, $\approx 0.2$. Note that this
discrepancy in the valence sector essentially originates from the
difference between non-standard flavor scheme (\ref{Par-Set-H})
and flavor symmetric polarized sea.

Finally, we summarize our generic model. Analogous as for PDFs in
Eqs.~(\ref{FjDelta0}) and (\ref{Dec-val/sea}), the odd $j$ (or
$-$) moments are entirely associated with valence quarks while the
even (or $+$) moments are given by the sum of sea and valence
quark contributions.  We plug the Mellin moments
(\ref{FjDelta0}) of our PDF ansatz (\ref{PDFmodel1}) into
Eq.~(\ref{Hpartialwave-1}) and obtain so the leading SO(3) PW
amplitudes
\begin{eqnarray}
\label{Mod-H}
&&\!\!\!\! \widetilde{H}_{jj}^{(3)+} = \widetilde{H}_{jj}^{(3){\rm val}}+
\widetilde{H}_{jj}^{(3){\rm sea}}\,,\qquad
\widetilde{H}_{jj}^{(3)-} =
\widetilde{H}_{jj}^{(3){\rm val}}\,, \quad\mbox{with}
\\
&&\!\!\!\! \widetilde{H}_{jj}^{(3)}(t)
=
g_A\, \frac{\left(1+j-\alpha\right)_p}{\left(1+j-\alpha(t)\right)_p}
\frac{
B(1+j-\alpha,1+\beta)- \widetilde{h}\, B(1+j-\alpha,1+\beta+\delta\beta)
}{
B(1-\alpha,\beta+1)- \widetilde{h}\, B(1-\alpha,1+\beta+\delta\beta)
}\,,
\nonumber
\end{eqnarray}
where the model parameters are specified in Eq.~(\ref{Par-Set-H}).

\subsection{Modelling $ \widetilde E_{jj}^{(3)}$}

Analogous as in the previous section, we build now a
generic model for the leading SO(3) PW amplitude $\widetilde
E_{j,J=j}(t)$. Thereby, we borrow us the $j$-dependence at
$t=0$ from the ansatz (\ref{Mod-H}) for $\widetilde
H_{jj}(t=0)$ with generic Regge intercept $\alpha=1/2$.
The new $\{\beta, \delta \beta, \widetilde e\}$
parameter set ($\widetilde h$ is replaced by $\widetilde e$, and to
lighten the notation, we use the same symbols $\beta$ and
$\delta \beta$ as for $\widetilde H$) will be specified  below.

At large $-t$ the helicity non-conserved moments $\widetilde
E_{jj}(t)$ should decrease by one power of $(-t)^{-1}$ faster than
$\widetilde H_{jj}(t)$, i.e., it should generically behave as
$1/(-t)^3$. In the heavy pion world, in which the $\pi_0/b_1$
trajectory is presumably degenerated with the $a_1/\rho_2$ one,
we would write the $t$-dependence of $\widetilde E_{jj}(t)$ as a
product of three monopoles,
$$
\frac{\left(j-\alpha_{\pi}\right)_3}{\left(j-\alpha_{\pi}(t)\right)_3} \,,
\quad (\mbox{where}\;\; j=J)\,,
$$
which are expressed in terms of the $\pi_0/b_1$ Regge trajectory
and its first  two daughters. In the real world  we choose  the
Regge trajectory (\ref{Reg-Tra-pi}), however, we assume that its
two daughters are now degenerated with those of the generic master
trajectory $\alpha(t)$. Finally, we introduce the partonic
decomposition of even and odd Mellin moments and we write our
ansatz for them as follows:
\begin{eqnarray}
\label{Mod-E0}
&&\!\!\!\! \widetilde{E}_{jj}^{(3)+} =
\widetilde{E}_{jj}^{(3),{\rm val}}+ \widetilde{E}_{jj}^{(3),{\rm sea}}\,,
\qquad
\widetilde{E}_{jj}^{(3)-} = \widetilde{E}_{jj}^{(3),{\rm val}}\,,
\\
&&\!\!\!\!
\widetilde{E}_{jj}^{(3)}(t) =
g_P \frac{\alpha^\prime_\pi\, \GeV^2}{j-\alpha_\pi(t)}
\frac{\left(1+j-\alpha\right)_2}{
\left(1+j-\alpha(t)\right)_2}
\frac{
B(1+j-\alpha,1+\beta)- \widetilde{e}\, B(1+j-\alpha,1+\beta+\delta\beta)
}{
B(1-\alpha,1+\beta)- \widetilde{e}\, B(1-\alpha,1+\beta+\delta\beta)
}\,,
\nonumber
\end{eqnarray}
where the generic and pion Regge trajectories are
specified in Eq.~(\ref{Reg-Par-pi}),
respectively. Our model for even $j$ is normalized at $j=J=0$,
\begin{equation}
\label{Mod-E-nor}
\widetilde{E}_{00}^{(3)+}(t) \equiv g_P(t)=
\frac{g_P\, \GeV^2}{m_\pi^2 -  t} \;\frac{\left(1-\alpha\right)_2}{
\left(1-\alpha(t)\right)_2}\,,
\quad g_P \equiv g^+_P = g^{\rm val}_P + g^{\rm sea}_P\,,
\end{equation}
by the pseudoscalar  coupling $g_P$. In our partonic
interpretation it contains a valence and sea quark part, where the
former one also determines the normalization of the  odd $j$ (or
$-$) GPD moments. The lowest moment (\ref{Mod-E-nor}) is nothing
but the pseudoscalar nucleon form factor. As designed, our
parameterization (\ref{Mod-E-nor}) possesses the pion pole, which
appears by means of Eq.~(\ref{Reg-Tra-pi}) as follows:
$$
\frac{\alpha^\prime_\pi}{j-\alpha_\pi(t)}\quad \stackrel{j=0}{\Rightarrow}
\quad
\frac{\alpha^\prime_\pi}{-\alpha_\pi - \alpha^\prime_\pi t} =
\frac{1}{m_\pi^2 -  t} \quad \mbox{with}\quad
m_\pi^2=-\frac{\alpha_\pi}{\alpha^\prime}\,.
$$

\begin{figure}[t!]
\centering
\includegraphics[width=17cm]{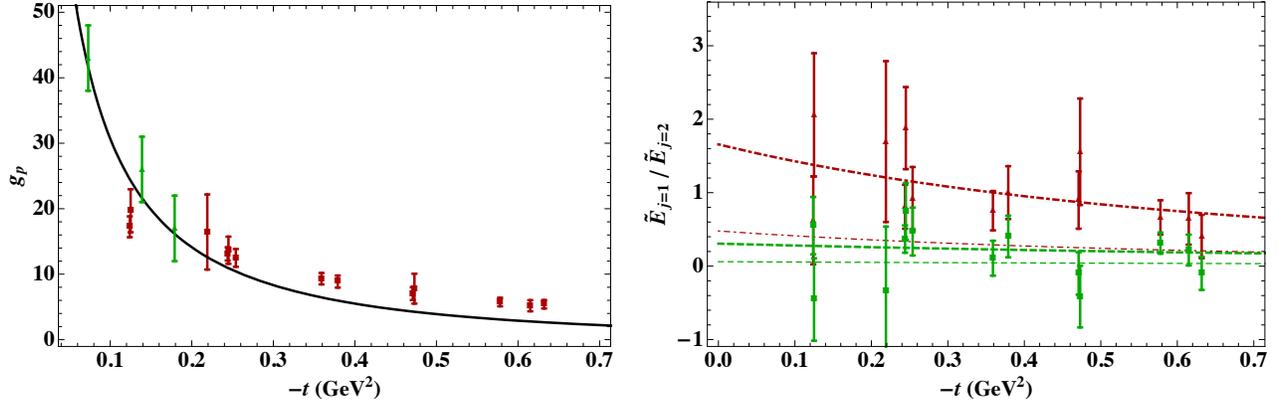}
\caption{\label{figparset-E}\small
In the left panel  the pseudo-scalar form factor is displayed:
generic model (\ref{Mod-E0}) prediction (solid curve),
experimental \cite{Choetal93} (triangles) and lattice  (squares)
data. In the right panel GPD model (\ref{Mod-E0}) predictions
within parameters (\ref{Par-SetE-e}) [thick]  and
(\ref{Par-Set-Ee1}) [thin] and lattice results for generalized
form factors are presented: $j=1$ (dot-dashed, triangles) and
$j=2$ (dashed, squares). Same scale specification and lattice data
as in Fig.~\ref{figparset}.
}
\end{figure}

Let us show that our generic GPD model satisfy the well-known low
energy constraints that arise from the hypothesis of a partial
conserved axial-vector current (PCAC). PCAC relates the
axial-vector and pseudo-scalar form factor, in our notation the
GPD combination (\ref{PWE-tHE}) for $j=0$, with the
pion-nucleon form factor:
\begin{eqnarray}
\label{Rel-PCAC}
\widetilde H^+_{00}(t) + \frac{t}{4 M^2} \widetilde E^+_{00}(t) =
\frac{f_\pi}{\sqrt{2}M}  \frac{m_\pi^2}{m_\pi^2-t} G_{\pi NN}(t)\,.
\end{eqnarray}
For the  pion-nucleon form factor  we adopt the popular monopole
parameterization
\begin{eqnarray}
\label{Def-GpiNN}
G_{\pi NN}(t) = \frac{G_{\pi NN}}{1-t/\Lambda^2}\,,
\quad G_{\pi NN} \approx 13\,,\;\; \Lambda \approx  0.8\, \GeV\,.
\end{eqnarray}
In the $t\to 0$ limit ($m_\pi >0$) the
Goldberger-Treiman relation shows up, which ties the axial-vector
coupling to the pion-nucleon coupling $g_A= f_\pi G_{\pi
NN}/\sqrt{2} M$ \cite{GolTre58}. In the chiral limit ($m_\pi \to
0$) the axial-current is conserved and the r.h.s.~of
Eq.~(\ref{Rel-PCAC}) vanishes. Our ansaetze (\ref{Mod-H}) and
(\ref{Mod-E0}) are so designed that  the resulting constraint
\begin{eqnarray}
\label{Rel-PCAC-1}
\lim_{m_\pi \to 0} \frac{-t}{4 M^2} \widetilde E^+_{00}(t)
=\widetilde H^+_{00}(t)
\quad\mbox{with}\quad
g_P=  4 g_A  M^2/\GeV^2
\end{eqnarray}
is satisfied within the well-known relation of pseudoscalar and
axial-vector coupling. If we now modify  the relation between
(partonic) pseudoscalar  and  axial-vector couplings by a small
$m_\pi^2$ proportional term,
\begin{eqnarray}
\label{Rel-gP-gA}
g^{\rm val/sea}_P=
g^{\rm val/sea}_A \frac{2-\alpha+m_\pi^2 \alpha^\prime}{2-\alpha}
4 M^2/\GeV^2\,,
\quad
g_A \equiv g_A^{\rm val}+g_A^{\rm sea} \approx 1.267\,,
\end{eqnarray}
and set in the pion-nucleon form factor (\ref{Def-GpiNN})
the monopole mass $\Lambda = \sqrt{(1-\alpha)/\alpha^\prime}$, our
model exactly satisfies the  PCAC relation (\ref{Rel-PCAC}).
Moreover, within the generic Regge parameters (\ref{Reg-Par-pi})
we find for the monopole mass the phenomenological value that is
quoted in Eq.~(\ref{Def-GpiNN}). Since we satisfy the PCAC
relation and have a realistic description of the axial-vector form
factor, see left panel in Fig.~\ref{figparset}, we also describe
in the left panel of Fig.~\ref{figparset-E} the $t$-dependence of
the pseudo-scalar form factor (\ref{Mod-E-nor}) measurements.

We specify now the remaining parameter set  $\{\beta,
\delta\beta,\widetilde e \}$ for both valence and sea quarks. As
in Sect.~\ref{subsec-ModH}, these parameters could be  partly read
off from large-$x$ counting rules. Unfortunately, to our best
knowledge,  for $\widetilde E$ GPD such counting rules are not
presented in the literature. We might expect that due to the
non-conserved target helicity a  $\widetilde E$ GPD  decreases by
two powers of $(1-x)$  faster at large $x$ than a $\widetilde H$
GPD. Our $\widetilde E$ GPD Mellin moments (\ref{Mod-E0}) behave
as $j^{-\beta-2}$ for $j\to \infty$ and, thus, in the limit $x\to
1$ the zero-skewness $\widetilde E$ GPD vanishes as
$(1-x)^{\beta+1}$.  For simplicity, we take this relative $(1-x)$
suppression factor in our  model, i.e., we take for the
$\widetilde E$ GPDs the same generic values for $\beta$ (and also
for $\delta\beta$) as for $\widetilde H$ GPDs, specified in
Eq.~(\ref{Par-Set-H}).

Now the $\widetilde e$ parameters can be used to tune the decrease
of the $\widetilde E$ GPD Mellin moments (\ref{Mod-E0}) with
growing $j$. To some extent they might be constrained by lattice
data for generalized form factors (set VI of
Ref.~\cite{Hagetal07}), shown in the right panel of
Fig.~\ref{figparset-E}. We use again our SU(6) symmetric valence
scenario (\ref{Par-Set-H}) within a flavor asymmetric sea and
evolve our moments, specified at $\mu^2=2\, \GeV^2$, to the scale
$\mu^2=4\, \GeV^2$ of Ref.~\cite{Hagetal07}. For $j=1$ we find
that $\widetilde e^{\rm val}$ is slightly larger than one
(triangles, thick dot-dashed curve). Since the $j=2$
generalized form factor data (squares) are compatible with zero and very
noisy, the sea quark parameter $\widetilde e^{\rm sea}$  is
essentially not constrained by data. We choose the parameter set
\begin{eqnarray}
\label{Par-SetE-e}
\widetilde{e}^{\rm val} \stackrel{\rm Lat}{=} 1.1\,,
\quad
\widetilde{e}^{\rm sea}\stackrel{\rm gen}{=}1\,,
\end{eqnarray}
which is fully compatible with lattice data (thick curves).
Below we will also use the set
\begin{eqnarray}
\label{Par-Set-Ee1}
{\widetilde e}^{\rm val}={\widetilde e}^{\rm sea} =0\,,
\end{eqnarray}
which is, however, disfavored by the lattice measurement of the
$j=1$ generalized form factor, see thin dot-dashed curve.

Finally, we summarize our  $\widetilde E_{jj}$ GPD model.  It
arises from plugging the PCAC relation (\ref{Rel-gP-gA}) into the
ansatz (\ref{Mod-E0}):
\begin{eqnarray}
\label{Mod-E}
&&\!\!\!\! \widetilde{E}_{jj}^{(3)+} =
\widetilde{E}_{jj}^{(3),{\rm val}}+ \widetilde{E}_{jj}^{(3),{\rm sea}}\,,
\qquad
\widetilde{E}_{jj}^{(3)-} = \widetilde{E}_{jj}^{(3),{\rm val}}\,,
\quad\mbox{with}
\\
&&\!\!\!\!
\widetilde{E}_{jj}^{(3)}(t) =
\frac{4 g_A \alpha^\prime\, M^2(2-\alpha- \alpha_\pi)\left(1+j-\alpha\right)_2
}{
(2-\alpha)(j-\alpha_\pi(t))\left(1+j-\alpha(t)\right)_2}
\frac{
B(1+j-\alpha,1+\beta)- \widetilde{e}\, B(1+j-\alpha,1+\beta+\delta\beta)
}{
B(1-\alpha,1+\beta)- \widetilde{e}\, B(1-\alpha,1+\beta+\delta\beta)
}\,,
\nonumber
\end{eqnarray}
the SU(6) symmetric valence scenario, cf.~Eq.~(\ref{Par-Set-H}),
generic Regge trajectories (\ref{Reg-Par-pi}), and by the $\widetilde e$
parameters (\ref{Par-SetE-e}) or, alternatively,
(\ref{Par-Set-Ee1}).

\subsection{The properties of a minimalist GPD model}
\label{subsec-GPD-Pro}

Let us shortly recall that at LO accuracy and for (nearly) fixed
photon virtuality $\cQ^2$ the GPD on the cross-over line $\eta=x$
entirely determines the `CFFs' \cite{Ter05}, see also
Ref.~\cite{KumMuePas08} for a phenomenological discussion to
reveal this GPD from experimental measurements. From
Eq.~(\ref{Def-CFF}) it follows that the imaginary part
\begin{eqnarray}
\label{GPDs2Im}
\Im {\rm m}
\left\{
{
\widetilde{{\cal H}}
\atop
\widetilde{{\cal E}}
}
\right\}(\xi,t,\cQ^2)
\stackrel{\rm LO}{=} \pi
\left\{
{
\widetilde{H}
\atop
\widetilde{E}
}
\right\}(x=\xi,\eta=\xi,t,\cQ^2)
\end{eqnarray}
of a `CFF' is directly expressed by the GPD on the cross-over
line. Moreover, for any GPD, i.e., this function has a double
distribution representation, the unsubtracted `dispersion
relations'
\begin{eqnarray}
\label{GPDs2Re+}
\Re {\rm e}
\left\{
{
\widetilde{{\cal H}}^+
\atop
\widetilde{{\cal E}}^+
}
\right\}(\xi,t,\cQ^2)
&\!\!\!\stackrel{\rm LO}{=}\!\!\!&
{\rm PV}\!\int_0^1\! dx\, \frac{2 \xi}{\xi^2-x^2}
 \left\{
 {
\widetilde{H}^+
\atop
\widetilde{ E}^+
}
\right\}(x,\eta=x,t,\cQ^2)\,,
\\
\label{GPDs2Re-}
\Re {\rm e}
\left\{
{
\widetilde{{\cal H}}^-
\atop
\widetilde{{\cal E}}^-
}
\right\}(\xi,t,\cQ^2)
&\!\!\!\stackrel{\rm LO}{=}\!\!\!&
{\rm PV}\!\int_0^1\! dx\, \frac{2 x}{\xi^2-x^2}
 \left\{
{
\widetilde{ H}^-
\atop
\widetilde{ E}^-
}
\right\}(x,\eta=x,t,\cQ^2)\,,
\end{eqnarray}
offer an alternative way to evaluate the real part of CFFs from
the GPDs on the cross-over line. To spell it out clearly, these
`dispersion relations' arise from the physical ones in the leading
twist-two approximation, i.e., the spectral function and its
support $|\xi| \le \xi_{\rm cut/pol} =1$ change
\cite{KumMuePas07}, and they are in one-to-one correspondence with
the LO approximation of CFFs (\ref{Def-CFF}).

Only the skewness dependence of the GPD on the cross-over line is
relevant for the LO description of experimental data. To quantify
this dependence,  we define the $r$-ratio
\begin{eqnarray}
\label{Def-Ske-Rat}
r(x,t,\cQ^2|{F}) = \frac{ F(x,\eta=x,t,\cQ^2)}{ F(x,\eta=0,t,\cQ^2)} \,,
\qquad {F} =\{{\widetilde H},{\widetilde E}\}
\end{eqnarray}
of the GPD on the  cross-over line and on the line $\eta=0$.
If this ratio is one, we call it zero skewness effect. Note that
the $r$-ratio is in a LO interpretation of DIS and hard
exclusive electroproduction to some extent `measurable' for ${H}$ and
${\widetilde H}$ type GPDs.

To build a model for the full ${\widetilde H}^{(3)\pm}$ and
${\widetilde E}^{(3)\pm}$ GPDs, we relate them by a `holographic
principle' to their forward analog. In the following we prefer
simplicity and take the {\em minimalist} `holographic principle'
in conformal Mellin space, see explanations in
Ref.~\cite{KumMue09}.  Namely, we equate the conformal GPD moments
\begin{eqnarray}
\label{GPD-Mod-Min}
\widetilde{H}_j(\eta,t) =
\widetilde{H}_{jj}(t)\, \hat{d}^{j+1}_{0,1}(\eta)
\quad\mbox{and}\quad
\widetilde{E}_j(\eta,t) =
\widetilde{E}_{jj}(t)\, \hat{d}^{j}_{0,0}(\eta)\,,
\end{eqnarray}
with the leading SO(3) PWs $J=j+1$ and $J=j$, specified in
Eqs.~(\ref{PW01}) and (\ref{PW00}), respectively. The price of
simplicity is that the GPD on the cross-over line is rigid. Our
leading SO(3)-PW model (\ref{GPD-Mod-Min}) can be equivalently set
up in  a minimalist ``dual'' parameterization or the Shuvaev
transform \cite{Shu99} (where we understand that this name already
implies the necessary restriction of the GPD transform
\cite{ShuBieMarRys99}). The corresponding GPD and the double
distribution have so far not been analytically evaluated, for other
model examples see \cite{MueSch05,PolSem08}. Numerically, the GPD can
be evaluated within a Mellin Barnes integral \cite{MueSch05}.

We are here only interested on the phenomenological GPD aspect and
need only the GPD on the cross-over line. Our specific model
assumptions imply a characteristic small-$x$ and large-$x$
behavior, which can be read off from the imaginary part of the
Mellin-Barnes integral (\ref{FFMellinmoments-+}) by means of
Eq.~(\ref{GPDs2Im}), where we set $\xi=x$. In the following we are
not concerned about evolution and set $\cQ^2=\cQ^2_0$ in
Eq.~(\ref{FFMellinmoments-+}).

The small-$x$ behavior is determined by the leading Regge pole,
which can be picked up by a shift of the integration path to the
left hand side. One easily finds a partial factorization of
$t$-dependence and a specific value for the $r$-ratio
\cite{ShuBieMarRys99},
\begin{eqnarray}
\label{GPD-beh-smax}
\lim_{x\to 0} r(x,t|{\widetilde H}) =
\frac{ 2^{\alpha(t)} \Gamma(3/2+\alpha(t))
}{
\Gamma(3/2)\Gamma(2+\alpha(t))}\,,
\quad
\lim_{x\to 0} r(x,t|{\widetilde E}) =
\frac{ 2^{1+\alpha_\pi(t)} \Gamma(5/2+\alpha_\pi(t))
}{
\Gamma(3/2)\Gamma(3+\alpha_\pi(t))}\,.
\end{eqnarray}
Since this is nothing but the Clebsch-Gordan coefficient in the
conformal PW expansion (\ref{FFMellinmoments-+}), we call it for $t=0$
conformal ratio%
\footnote{The name is also motivated by the fact that a minimalist GPD
model with leading Regge behavior in the small $x$-region and for
$t=0$ is equivalent to a conformal GPD model. Since effective
Regge behavior is commonly present in a realistic GPD model, we
name the ratio for shortness conformal.} \cite{KumMue09}. This  ratio is
only a function of the leading trajectory $\alpha(t)$ and it is
valid for a restricted $\alpha(t)$ interval, i.e., $-t$ is
restricted by an upper bound. For  $\alpha(t) >0$ and
$\alpha_\pi(t) >-1$ our models possess a positive skewness effect $r
> 1$.

The large-$x$ behavior of a minimalist  GPD on the cross-over
line, can be derived from the large-$j$ behavior of the
Mellin-Barnes integrand (\ref{FFMellinmoments-+}) taken at large
$x$. It arises from the Clebsch-Gordon coefficient, the
skewness-zero GPD moments, and the behavior of SO(3) partial waves
$$ \hat d^{j+\lambda}_{0,\lambda}(x)\sim
\frac{\Gamma(3/2)\Gamma(3+j)}{\Gamma(5/2+j)}  (x/2)^{j+1}
\; 2^{2-3\lambda} (3+2j)^\lambda \;e^{j \sqrt{1-x^2}}
\quad
\mbox{with}\quad\lambda=\{0,1\}.
$$
Here we utilized the definitions (\ref{PW01}) and (\ref{PW00}) in
terms of hypergeometric functions and their large $j$ asymptotic
for $x\to 1$ \cite{Luk169}. Combining all together, and
considering the Mellin-Barnes integral as a Fourier transform with
conjugate variable $\sqrt{1-x^2} \sim \sqrt{2}\sqrt{1-x}$ , we
find that our minimalist GPDs behave at $x \to 1$ as
\begin{eqnarray}
\label{GPD-beh-larx}
{\widetilde H}(x,x,t) &\!\!\!\sim\!\!\!&
(1-x)^{\frac{\beta-1}{2}}\quad
\stackrel{\beta =3}{\Rightarrow} \quad  \sim (1-x)^{1}\,,
\\
{\widetilde E}(x,x,t) &\!\!\!\sim\!\!\!&
(1-x)^{\frac{\beta+1}{2}}\quad \stackrel{\beta =3}{\Rightarrow}
\quad  \sim (1-x)^{2}\,.
\nonumber
\end{eqnarray}
As required, our findings are consistent  with the
large-$x$ result in the `dual' model%
\footnote{For $H$ GPD one might utilize the SO(3)-PWs
$d^{J}_{0,0}$ and so the power is  $\beta/2$ in agreement with
Ref.~\cite{Pol07,PolSem08}. Within SO(3)-PWs $d^{J}_{0,1}$
polynomiality is  not completed  and we obtain  $(\beta-1)/2=1$
for $\beta=3$, which is consistent with the large $x$-behavior  in
Radyushkin's double distribution ansatz with $b=1$\cite{Rad98}. }
\cite{Pol07}. The behavior (\ref{GPD-beh-larx}) also coincides
with the large-$x$ counting as it has been derived for the  $H$
and $E$ GPDs from diagrammatic considerations \cite{Yua03}. Hence,
compared to the zero-skewness GPD, our GPDs at the cross-over line
possess a large enhancement effect. We also note that in our
models within linear Regge trajectories the $t$-dependence dies
out at large $x$ with $\sqrt{1-x}$. For fixed $t$ the skewness
ratio in the large $x$-region can be written  as
\begin{eqnarray}
\label{GPD-beh-larx-r}
\lim_{x\to 1} r(x,t|{F}) \sim
(1-x)^{-\frac{\beta+1}{2}}
\frac{F^{\rm red}(\sqrt{2}\sqrt{1-x}\,t)}{F^{\rm red}((1-x)t)}\,,
\end{eqnarray}
where $F^{\rm red}$ is a reduced GPD that encodes the
$t$-dependence for both $\eta=0$ and $\eta=x$ cases. From the
point of view of light-cone wave function modelling,
\cite{Bur04,Bur07}, double distribution models
\cite{MukMusPauRad02,TibDetMil04,HwaMue07}, and diagrammatical
counting rules \cite{Yua03} one expects that the $t$-dependence
dies out faster for $x\to 1$. It is here far beyond the scope to
fix the presumable unrealistic feature in our models, e.g., by
employing slightly non-linear Regge trajectories.

\begin{figure}[htb]
\centering
\vspace{4mm}
\includegraphics[width=17cm]{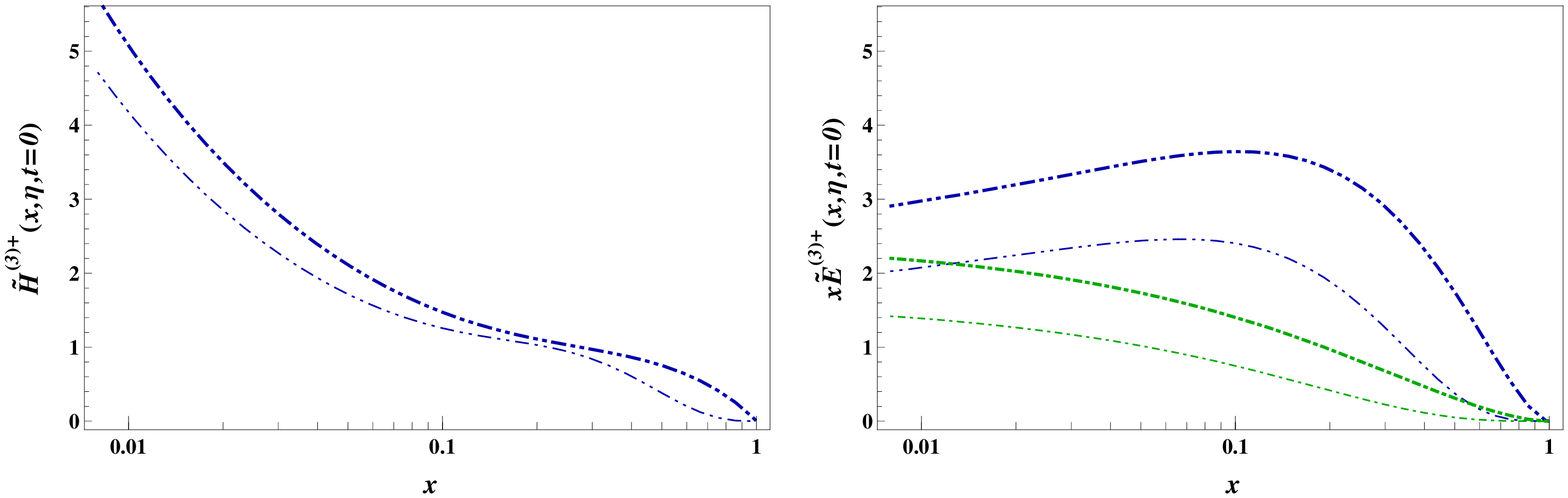}
\vspace{-4mm} \caption{\label{Fig-GPD-1}  \small
The charge parity-even ${\widetilde
H}^{(3)+}(x,x,t=0)$ GPD (left)  and scaled $x {\widetilde
E}^{(3)+}(x,x,t=0)$ GPD  (right) with
parameter sets (\ref{Par-Set-H}) and (\ref{Par-Set-Ee1})  are
shown as dot-dot-dashed curves and for parameters
(\ref{Par-SetE-e}) as dot-dashed. The corresponding GPDs at
$\eta=0$ are displayed as thin curves.}
\end{figure}
For $t=0$  we illustrate the skewness effect  in
Fig.~\ref{Fig-GPD-1}, where we show GPDs on the lines $\eta=x$ and
$\eta=0$ as thick and thin curves, respectively. More
specifically, we plot the dominant charge parity-even GPD models
${\widetilde H}^{(3)+}$   (left) and $x {\widetilde E}^{(3)+}$
(right), where the latter is rescaled by $x$. Thereby, the
${\widetilde H}$ (\ref{Mod-H}) and ${\widetilde E}$ (\ref{Mod-E})
GPDs are specified by the parameter sets (\ref{Par-Set-H}) and
(\ref{Par-SetE-e}), shown as dot-dot-dashed curve, and
alternatively for the $\widetilde E$ GPD within the parameters
(\ref{Par-Set-Ee1}) as dot-dashed curves. As one realizes in the
right panel the $\widetilde E$ GPD models are qualitatively
different, however, they approach each other in the small $x$-region.
The characteristic small-$x$ enhancement of a minimalist GPD is
controlled by the conformal ratio (\ref{GPD-beh-smax}). For our
$\widetilde H$ and $\widetilde E$ GPDs  it is given by $r \approx
1.2$ and $r \approx 1.5$, respectively. The large-$x$ enhancement
(\ref{GPD-beh-larx}) is clearly a big effect for the
dot-dot-dashed models. We also observe that, compared to its
forward analog, our minimalist GPDs on the cross-over line are
enhanced in the valence region. Note that our $\widetilde E$ GPD
on the cross-over line is always positive, while in the $\chi$QSM
estimate it can be even negative \cite{PenPolGoe99}.

Let us now point out the particularities of the charge parity-even
$\widetilde{E}^{(3)+}$ GPD. This GPD steeply rises as
$x^{-1-\alpha_\pi(t)}$ with decreasing $x$, see the flat behavior
of the scaled $x \widetilde{E}$ GPD in the right panel of
Fig.~\ref{Fig-GPD-1}. Since the small intercept $\alpha_\pi = -
m_\pi^2 \alpha^\prime$ is negative, it is guaranteed that for $t
\le 0$ the lowest Mellin moment of this GPD exists and that an
unsubtracted dispersion relation (\ref{GPDs2Re+}) is sufficient.
Because of this singular behavior at $x=0$, the  charge
parity-even `CFF' (\ref{GPDs2Re+}) possesses a large real part
that can be loosely viewed as a $1/\xi$ proportional constant. To
isolate this term, we introduce an over subtraction in the
dispersion relation (\ref{GPDs2Re+}) at the point $\xi = \infty$:
\begin{eqnarray}
\label{DisRel-new}
\Re {\rm e}  \widetilde{{\cal E}}^{(3)+}(\xi,t) \stackrel{\rm LO}{=}
\frac{1}{\xi} {\rm PV}\! \int_0^1\! dx\, \frac{2 x^2}{\xi^2-x^2}
\widetilde{{E}}^{(3)+}(x,x,t)\, +
\frac{1}{\xi} {C}_{\cal E}(t)\,, \quad
{C}_{\cal E}(t)=2\int_0^1\! dx\,  \widetilde{{\cal E}}^{(3)+}(x,x,t)\,.
\end{eqnarray}
The subtraction constant in this new `dispersion relation' is
evaluated from the GPD on the cross-over line and, thus, it
encodes the skewness effect. In the vicinity of $t=m_\pi^2$, we
find by means of the Mellin-Barnes integral
(\ref{FFMellinmoments-+}) the pion pole:
\begin{eqnarray}
\label{PioPol-MinGPD}
{C}_{\cal E}^{\rm Min}(t) \stackrel{\rm LO}{=}
\frac{3}{2} \frac{g_P}{m_\pi^2 -  t} +
\mbox{non pion-pole contributions}\,.
\end{eqnarray}
This is nothing but the result in the parameterization of
Ref.~\cite{ManPilRad98,FraPobPolStr99},
\begin{eqnarray}
\label{Def-PioDom}
\widetilde {\cal E}_{\pi-{\rm pole}}^+(\xi,t) =
\frac{1}{\xi} {C}^{\rm Min}_{\pi-{\rm pole}}(t)\,,
\quad
{C}^{\rm Min}_{\pi-{\rm pole}}(t)=
\int_{-1}^{1}\!dx\,\frac{1}{\xi-x}
\frac{g_P}{m_\pi^2 -  t} \frac{\theta(|x| \le \xi) }{2\, f_\pi}\;
\phi^{\rm asy}_\pi\!\left(\!\frac{\xi-x}{2\xi}\!\right)\,,
\end{eqnarray}
where the residue is provided by the {\em asymptotic} pion DA
\begin{eqnarray}
\label{DA-pion}
\phi^{\rm asy}_\pi(u)= f_\pi 6 u(1-u).
\end{eqnarray}

Few comments are in order, which are partially adopted from
discussions of parity-even GPDs and the so-called $D$-term
\cite{PolWei99} in Refs.~\cite{KumMuePas07,KumMuePas08}:

\vspace{2mm} \noindent {\em i}.\phantom{ii}~In the
parameterization of Ref.~\cite{ManPilRad98,FraPobPolStr99} the
pion-pole lives only in the central GPD region and it is decoupled
from the smooth GPD part. This might be the natural answer that
arises from a dynamical model calculation, in particular, if the
chiral limit is taken. In such a model the cross talk of imaginary
and real part is interrupted. On the other hand we can consider
the chiral limit of the subtracted dispersion relation
(\ref{DisRel-new}). Although the intercept is now $\alpha_\pi=-1$,
the integral still exists for $t < 0$ and for $t\ge 0$ the answer
can be obtained by an analytic regularization of the integral.

\vspace{1mm}\noindent
{\em ii}.\phantom{i}~One might also reverse the
logic, and utilize the subtraction constant (\ref{DisRel-new}) as a sum rule
that constraints the GPD on the cross-over line.

\vspace{1mm} \noindent
{\em iii}.~The residue of the pion pole in the minimalist GPD model, see
Eq.~(\ref{PioPol-MinGPD}), is equal to that one in
Eq.~(\ref{Def-PioDom}), obtained from the asymptotic DA.  This is
{\em not} accidental and it can be viewed from a more general
perspective, reflecting the crossing relation between $t$-
and $s$-channel. Namely, a family of asymptotic generalized DAs,
labelled by the $t$-channel angular momentum $J$ is the crossing
analog of the minimalist GPD.
\vspace{2mm}

\begin{figure}[thb]
\centering
\vspace{4mm}
\includegraphics[width=17cm]{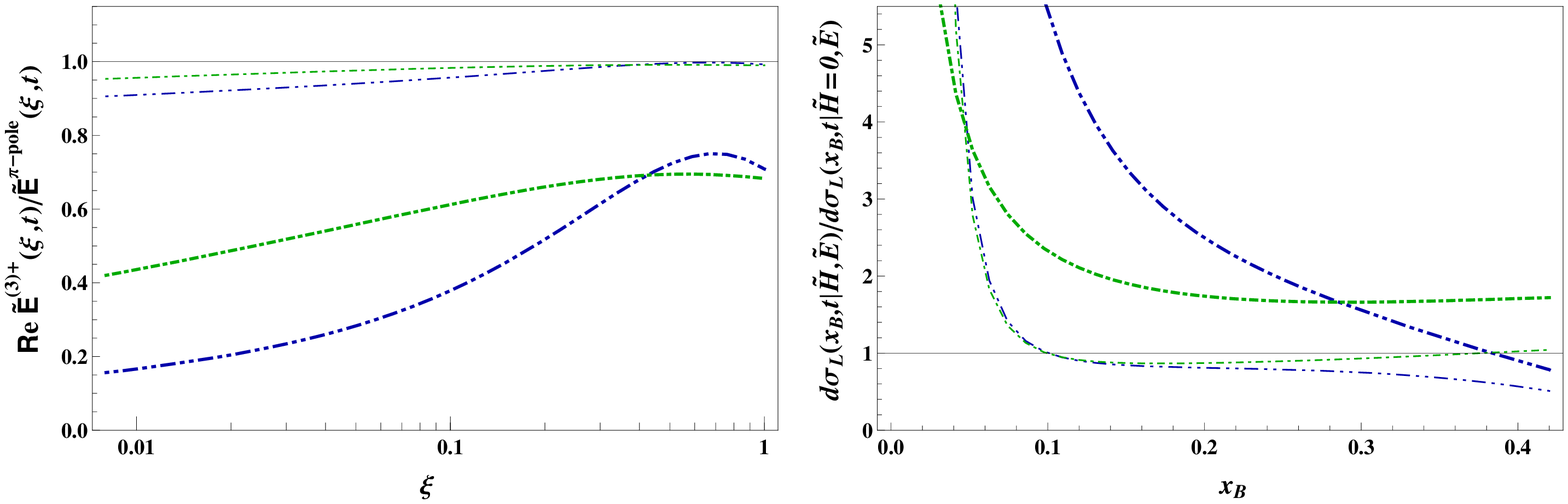}
\vspace{-4mm} \caption{\label{Fig-GPD-2}\small Left panel:
$\Re{\rm e}{\widetilde {\cal E}}^+(\xi,t)/{\widetilde {\cal
E}}^+_{\pi-{\rm pole}}(\xi,t)$ ratio, cf.~Eq.~(\ref{Def-PioDom}),
versus $\xi$ for fixed $t=0$ (thin)  and $t=-0.2\,\GeV^2$ (thick).
Right panel:  cross section ratio
$d\sigma_L(\Bx,t^\prime|\widetilde H,\widetilde
E)/d\sigma_L(\Bx,t^\prime|\widetilde H\equiv0,\widetilde E)$
versus $\Bx$ for $t^\prime =0$ (thin) and $t^\prime=-0.2\,\GeV^2$
(thick) with ${\cal Q}^2= 2\,\GeV^2$. The dot-dot-dashed and
dot-dashed curves refer to the parameterizations
(\ref{Par-SetE-e}) and (\ref{Par-Set-Ee1}), respectively.}
\end{figure}
In Fig.~\ref{Fig-GPD-2} we illustrate the role of the pion pole
contribution for GPD modelling. In the left panel we show the
ratio of the real part of our `CFF' model to that one of
Ref.~\cite{VanGuiGui99}, entirely given by the pion pole
contribution (\ref{Def-PioDom}), versus $\xi$. For $t=0$ the ratio
is shown as thin curves and as expected it is about one.  It
slightly gets smaller with decreasing $\xi$, where the  comparison
of the dot-dot-dashed (\ref{Par-SetE-e}) and dot-dashed
(\ref{Par-Set-Ee1}) curves reveals that the parameterization of
the GPD on the cross-over line has some influence. For
non-vanishing $-t$ the difference is more pronounced. Since we
neglected the residual $t$-dependence in the pion pole
(\ref{Def-PioDom}), the  ratio considerably decreases for
$-t=0.2\,\GeV^2$ (thick curves). In the right panel we display the
ratio of the unpolarized cross sections (\ref{XsectionFinal})  for
the full $\widetilde H$ and $\widetilde E$ GPD model
(\ref{Mod-H},\ref{Mod-E}) to that one in which $\widetilde H$ is
neglected. Thereby, we set $t=t_{\rm min} +t^\prime$. Clearly,
only for $t^\prime=0$ and in some limited $\Bx$ interval this
ratio is around one and so the $\widetilde E$ dominance is
justified. We conclude that in general kinematics, even in the
valence quark region for smaller values of $-t^\prime$, the hard
exclusive $\pi^+$ electroproduction cross section might {\em not}
be dominated by the pion pole (\ref{Def-PioDom}).

\section{Measurements versus formalism and minimalist  model}
\label{Sec-ExpResGPD}

We like now to confront our minimalist GPD models with the exclusive
$\pi^+$ electroproduction cross section \cite{Airetal07} and  single
transverse proton spin asymmetry \cite{Hri08} measurements from
the HERMES collaboration and the longitudinal photoproduction
cross section, measured in Hall C at JLAB,
\cite{Horetal07,Bloetal08}. However, first we  will define the
observables that can be related to the collinear factorization
approach. We like then to comment on the theoretical problems of
the collinear factorization approach and we suggest a simple
prescription to cure its LO approximation. We finally rely on this
prescription and compare our ad hoc GPD models and a tuned one
with experimental data.

It is common to convert the electroproduction cross section to the
virtual photoproduction one by means of the Hand convention for
the photon flux factor $\Gamma_V$:
\begin{eqnarray}
\label{XXcon-Exp}
\frac{d\sigma^{\gamma^*p\rightarrow n \pi^+}}{dt} =
\frac{1}{\Gamma_V(\Bx,\cQ)}  \int_{0}^{2\pi}\! d\varphi\,
\frac{d\sigma^{\pi^+}}{d{\cQ}^2  d\Bx dt d\phi}
\,,\quad
\Gamma_V =
\frac{\alpha_{\rm em}}{2 \pi}
\frac{y^2}{1-\veps} \frac{1-\Bx}{\Bx \cQ^2}\,.
\end{eqnarray}
The  photoproduction  cross section can be
separated in the transverse and longitudinal part
\begin{eqnarray}
\frac{d\sigma^{\gamma^*p\rightarrow n \pi^+}}{dt} =
\frac{d\sigma_T^{\gamma^*p\rightarrow n \pi^+}}{dt} + \veps
\frac{d\sigma_L^{\gamma^*p\rightarrow n \pi^+}}{dt}\,.
\end{eqnarray}
Utilizing the Hand convention (\ref{XXcon-Exp}), we can read off
the perturbative prediction for the longitudinal polarized
photoproduction cross section from Eq.~(\ref{XsectionFinal}):
\begin{eqnarray}
\label{phoXsectionFinal}
\frac{d\sigma_L^{\gamma^*p\rightarrow n \pi^+}}{dt} &\!\!\!
\stackrel{{\rm Tw}-2}{=} \!\!\! &
\frac{ 4 \pi^2\alpha_{\rm em}}{\sqrt{1+\epsilon^2} {\cal Q}^6}
\frac{\Bx^2}{1-\Bx}
\Bigg\{
(1-\xi^2) |\widetilde {\mathrm H}_{\pi^+}|^2
-\frac{t}{4 M^2} |\xi\widetilde {\mathrm E}_{\pi^+}|^2
- 2 \xi^2 \Re {\rm e}\, \widetilde {\mathrm H}_{\pi^+}
\widetilde {\mathrm E}^{\ast}_{\pi^+}
\Bigg\}\Bigg|_{\xi \to \frac{\Bx}{2-\Bx}}.
\nonumber\\
\end{eqnarray}

If the extraction of the longitudinal photoproduction cross
section  is experimentally not reachable, a comparison of GPD
model predictions for the longitudinal cross section
(\ref{phoXsectionFinal}) and electroproduction cross section
measurements requires a quantitative understanding of its
transverse part, too. Utilizing the $R$ ratio, we might express
the measured differential photoproduction cross section
(\ref{XXcon-Exp}) to its longitudinal part
(\ref{phoXsectionFinal}) as follows:
\begin{eqnarray}
\label{Pre-CroSec}
\frac{d\sigma^{\gamma^*p\rightarrow n \pi^+}}{dt} =
\left[1 + \frac{1}{\veps(\Bx,\cQ^2) R(\Bx, t,\cQ^2)}
\right] \veps(\Bx,\cQ^2)
\frac{d\sigma_L^{\gamma^*p\rightarrow n \pi^+}}{dt}\,, \quad
R =\frac{d\sigma_L^{\gamma^*p\rightarrow n \pi^+}}{
d\sigma_T^{\gamma^*p\rightarrow n \pi^+}}\,.
\end{eqnarray}
Note that the factorization theorem \cite{ColFraStr96} tells us
only that the inverse ratio $1/R$ formally vanishes with
$1/\cQ^2$, however, it does not state that it is numerically small
at accessible photon virtualities. Rather it should be expected
that $1/R$ is numerically enhanced by non-factorizable
contributions, which might also include rather large logarithmical
modification of its expected $1/\cQ^2$ fall-off;  for an example
see the perturbative study to the proton Pauli form factor in
Ref.~\cite{BelJiYua03}. Indeed, measurements in Hall C at JLAB
reveal that $1/R$ is rather sizeable for $\cQ^2 < 4\, \GeV^2$ and
$W \approx 2\, \GeV $ \cite{Horetal07,Bloetal08}.

Unfortunately, to our best knowledge, also a model understanding of
the $1/R$ ratio is at present not fully reached. For instance, the
Regge-inspired model \cite{GuiLagVan97} describes at such low $W$
the longitudinal and the real transverse photoproduction cross
section measurements, however, underestimates the virtual
transverse one. On the other hand it is in fair agreement with the
HERMES measurements for the virtual photoproduction cross section
\cite{Airetal07}. A refined version of this model, tuned to
low energy data from JLAB \cite{KasMurMos08},  does describe most
of  HERMES data, too, however not those at large photon
virtualities and relative low energy \cite{KasMos09}. In the
models the pion pole is reggeized and it essentially contributes
only to the longitudinal cross section, which for $|t^\prime| \le
0.5\,\GeV^2 $ is dominant. Hence, for smaller $-t^\prime$ values
both models suggest that we can set $1/R$ to zero.

Only the $\sin(\phi-\phi_S)$ harmonic  of the single transverse
proton spin asymmetry
\begin{eqnarray}
\frac{d^{\uparrow}\sigma - d^{\downarrow}\sigma
}{
d^{\uparrow}\sigma + d^{\downarrow}\sigma}
= A_{\rm UT}^{\sin(\phi-\phi_S)} \sin(\phi-\phi_S) +
\mbox{other harmonics}\,,
\end{eqnarray}
is counted as a further twist-two observable
\cite{FraPobPolStr99}, while other harmonics are at least formally
suppressed by $1/\cQ$. In the notation of Ref.~\cite{DieSap05}
the asymmetry is written in terms of so-called polarized
photoabsorption cross sections $\sigma_{mn}^{ij}$:
\begin{eqnarray}
\label{Def-Asy-DieSap}
A_{\rm UT}^{\sin(\phi-\phi_S)} =-\frac{
\Im{\rm m}\left(\veps\, \sigma_{00}^{+-} + \sigma_{++}^{+-}\right)
}{ \veps\,\sigma_{00}^{++} + \frac{1}{2}\left(
\sigma_{++}^{++}+ \sigma_{++}^{--}\right) }\,,
\end{eqnarray}
where the subscripts (superscripts) refer to the photon (target)
helicity. Both terms $\sigma_{++}^{+-}$   and $\sigma_{++}^{++}+
\sigma_{++}^{--} \propto d\sigma_T$ arise from the exchange of a
transverse polarized photon and, thus, are  counted as  $1/\cQ^2$
power suppressed non-factorizable contributions.  We neglect both
of them and from the longitudinal cross section
(\ref{XsectionFinal}) we read off the twist-two approximation for
the asymmetry
\begin{eqnarray}
\label{Def-AUT}
A_{\rm UT}^{\sin(\phi-\phi_S)} \stackrel{\rm Tw-2}{=}
\frac{
-2
\sqrt{\frac{t-t_{\rm min}}{t_{\rm min}}}\,  \xi^2\
\Im {\rm m}\,
\widetilde {\mathrm H}_{\pi^+} \widetilde {\mathrm E}^{\ast}_{\pi^+}
}{
(1-\xi^2) |\widetilde {\mathrm H}_{\pi^+}|^2
-\frac{t}{4 M^2} |\xi\widetilde {\mathrm E}_{\pi^+}|^2
- 2 \xi^2 \Re {\rm e}\, \widetilde {\mathrm H}_{\pi^+}
\widetilde {\mathrm E}_{\pi^+}^{\ast}
}
\Bigg|_{\xi \to \frac{\Bx}{2-\Bx}}\,.
\end{eqnarray}
Note that the authors of Ref.~\cite{FraPobPolStr99} argued that
non-perturbative contaminations  mostly cancel in this asymmetry
and that a cancellation of perturbative corrections has been
numerically established within some GPD models in Refs.
\cite{BelMue01a,DieKug07}.

As discussed in Sect.~\ref{Sec-Int}, it is an intricate task to
judge on the collinear factorization approach, applied to the hard
exclusive meson electroproduction. We recall that the hard
amplitude is rather analogous as for the pion form factor, where
the momentum fractions of the incoming minimal pion Fock state are
replaced by those of the emitted and absorbed quarks
\cite{BelMue01a}, see Fig.~\ref{Factorization}b. Moreover, the
pion pole plays an important role at very small $-t$ and not too
small $\Bx$.  Thus, the theoretical descriptions of the hard
exclusive $\pi^+$ electroproduction and pion form factor have
similarities. The perturbative framework for the pion form factor
has been widely discussed in the literature, for critics see
Ref.~\cite{IsgLle89a,Rad91}, and various improvements were
suggested. One might use these recipes to improve the description
of hard exclusive meson electroproduction, too, e.g., going back
to the overlap representation of Drell and Yan in which the meson
wave functions contains transverse degrees of freedom
\cite{DreYan69}, modifying the factorization by incorporating the
Sudakov suppression \cite{LiSte92,JakKro93}, or a scale setting
prescription that reshuffles perturbative corrections to the
non-perturbative region of the coupling \cite{BroRJiPanRob97}.

A popular GPD code is called VGG plus ``power corrections''. Here
the ``power corrections'' result from a `hand-bag' calculation in
which transverse degrees of freedom (and only the pion pole in the
$\widetilde E$ GPD) are taken into account \cite{VanGuiGui99}. It
should be clear from the very beginning that these ``power
corrections'' can not be considered as autonomous contributions
and that their interplay with higher Fock state components or
so-called soft contributions is not obvious; see also for
illustration the light-cone sum rule analysis in
Ref.~\cite{BraKhoMau99}. In some sense this recipe is analog to
the Drell-Yan formula for the pion form factor and here the
conceptual problem of such an approach is obvious. Namely, the
Drell-Yan formula is the starting point and integrating out the
transverse degrees of freedom yields the collinear factorization
approach. This procedure generates perturbative corrections in the
leading twist-two sector \cite{BroLep80} and also power suppressed
contributions. Strictly spoken, even perturbative and power
suppressed contributions can not be autonomously considered as
they have a cross talk. Already, for some time it was observed
that the sign and magnitude of ``power corrections''
\cite{VanGuiGui99} and perturbative NLO corrections  in both DVCS
\cite{BelMueNieSch99,KumMuePas07} and hard exclusive meson
electroproduction \cite{BelMue01a} match. Although this
observation is not quantitatively understood, it reveals the
problem of double counting.   Hence, one can not take the notion
VGG plus  ``power corrections''  literally.

It would be of immense phenomenological interest to estimate the
size of perturbative and non-perturbative corrections to the LO
formulae (\ref{MesonProFunctions}), (\ref{LOamplitude}) of the
collinear factorization approach. Since not much is known about
power suppressed corrections, we will here assume that they can be
neglected. The biggest portion of the large perturbative NLO
corrections is proportional to the first coefficient of the
$\beta(\alpha_s)$ function \cite{BelMue01a}, which governs the
scale change of the running coupling (\ref{Def-alpha}). These
large corrections can be viewed as an indication for a break down
of perturbation theory. This calls for a resummation procedure,
which naturally induces its own ambiguity. Since the LO
approximation of the hard scattering amplitude (\ref{LOamplitude})
is proportional to the running coupling, a big portion of the NLO
corrections can be absorbed by changing the renormalization scale
$\mu_R$. The ambiguity we fix by the Brodsky-Lepage-Mackenzie
(BLM) scale setting prescription \cite{BroLepMac83} (for the
treatment of complex amplitudes see Ref.~\cite{BroLlaEst06}),
yielding
$$
\alpha_s\left(\mu^{\rm BLM}_R=\exp\{-f(\xi)/2\} \cQ\right)
\quad\mbox{with}\quad
e^{-f(\xi)} \lesssim  0.1\,.
$$
The resulting renormalization scale, i.e., the function $f(\xi)$,
depends on the GPD itself, however, it will be certainly driven
into the infrared region and so the coupling itself is now
plugged by non-perturbative physics.

The BLM scale in hard exclusive electroproduction is rather small
\cite{BelMue01a,AniPirSzyTerWal04}, similar as in the perturbative
description of the pion form factor
\cite{BroRJiPanRob97,MelNicPas02}. In the following we do not aim
for an ``optimal'' description of the hard exclusive $\pi^+$
production rather for a semi-quantitative understanding.
Therefore, we simply assume that the effective couplings, arising
from the BLM scale setting procedure and the freezing in the
infrared region, have in both processes the same value. Hence,
within a very low renormalization scale we can employ the LO
approximation for the (scaled) pion form factor,
\begin{eqnarray}
\label{Ref-ScaFF-pion}
\cQ^2 F_\pi(\cQ^2) \stackrel{\rm LO}{=}
8 \pi \alpha_s^{\rm eff} f_\pi^2\, {\cal I}^2_\pi(\cQ^2)\,,
\end{eqnarray}
to find the value of this effective coupling. As mentioned,  the
inverse moment $ {\cal I}_\pi $, defined in
Eq.~(\ref{Def-ConInt}), can be directly revealed at LO from the
photon-to-pion transition form factor. Its value, determined from
CLEO data \cite{Sav97}, would be slightly smaller than one,
see, e.g., Refs.~\cite{MusRad97,MikSte09} for comprehensive
discussions.  We set it here generically to one, i.e., ${\cal
I}_\pi(\cQ^2) =1$. This value also arises from  the asymptotic
pion DA (\ref{DA-pion}) however, also  pion DAs with a more
complex shape  might approximately possess the same value
\cite{BakMikSte01,BakMicSte05,MikSte09}. For the $\cQ^2$  scaled
pion form factor (\ref{Ref-ScaFF-pion}) we take the value $0.36$,
which is for $\cQ^2 > 1.5\,\GeV^2$ and within 1-$\sigma$ standard
deviation compatible with the model dependent extraction of the
form factor from the $\pi^+$ electroproduction \cite{Bloetal208},
and so we obtain the following value for the effective coupling:
\begin{eqnarray}
\label{Cou-Fro}
\alpha_s^{\rm eff} = 0.8\,.
\end{eqnarray}
Both the photon-to-pion transition and the elastic pion form
factors are described at LO. Note that also within the BLM setting
prescription sizeable NLO corrections, related to the Sudakov
suppression, affect the LO description of the pion form factor
\cite{BroRJiPanRob97}.

\begin{figure}[t]
\centering
\hspace{-5cm}
\includegraphics[width=12.5cm]{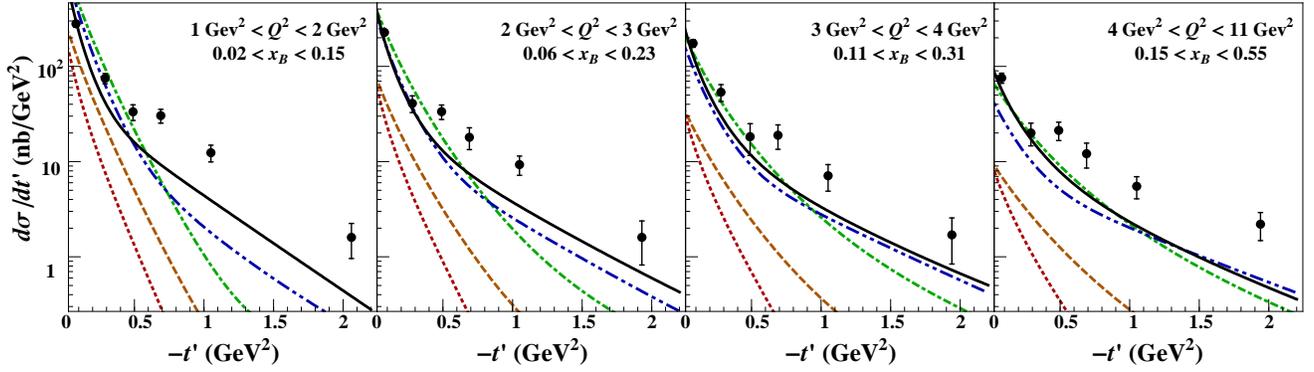}
\caption{\label{FigdXSec}\small  HERMES measurements \cite{Airetal07} of
the differential cross section
$d\sigma^{\gamma^*p\rightarrow n \pi^+}/dt'$  versus
minimalist GPD model predictions: only with pion pole
contribution (dotted), including $\widetilde H$ (dashed),  full
model with the setting (\ref{Par-Set-Ee1}) and frozen coupling
(dot-dashed), same as dot-dashed within parameters
(\ref{Par-SetE-e}) (dot-dot-dashed), and versus a tuned
GPD model (\ref{Par-Set-emp}) (solid).}
\end{figure}
\begin{figure}[t]
\includegraphics[width=17cm]{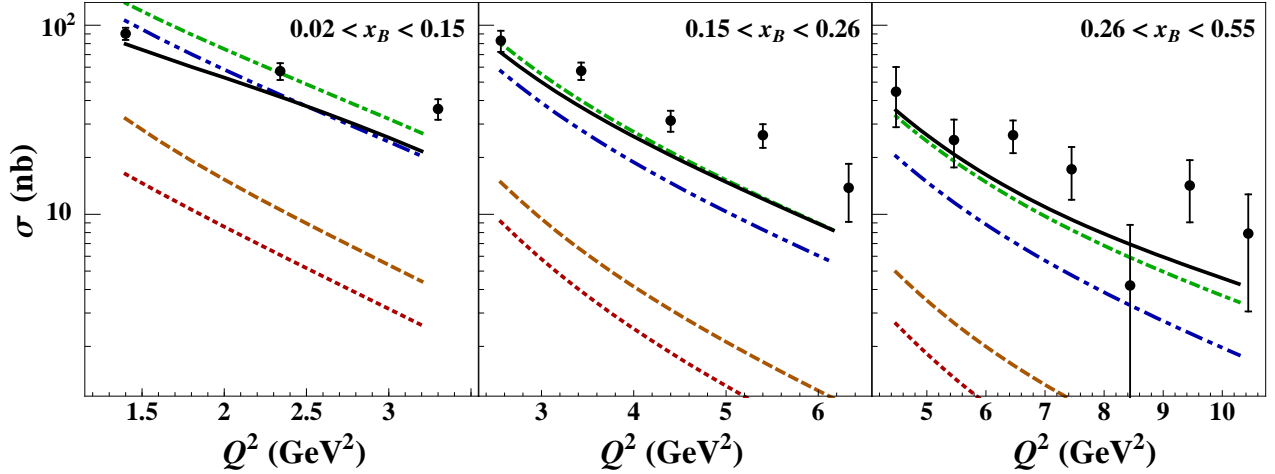}
\caption{\label{FigTotX} \small Total  cross section $\sigma^{\gamma^*p\rightarrow n
\pi^+}$ versus $\cQ^2$ measurements from HERMES \cite{Airetal07} are shown together with
a tuned (solid) and various  ad hoc GPD model estimates with $1/R=0$.
Same models as in Fig.~\ref{FigdXSec}.
}
\end{figure}
\begin{figure}[t]
\centering
\hspace{-3.5cm}
\includegraphics[width=13cm]{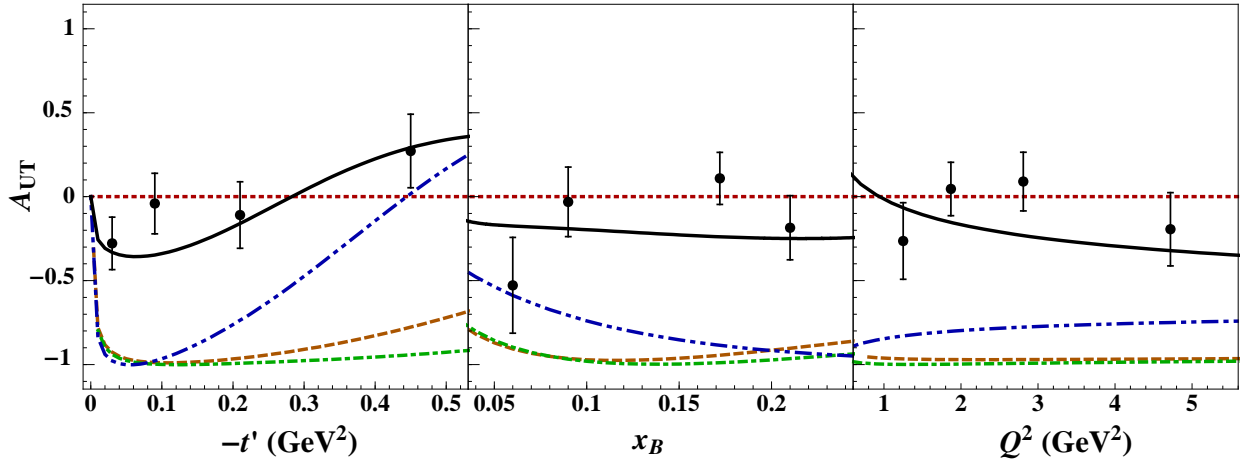}
\caption{\label{FigTPSA}\small
Preliminary  HERMES measurements
\cite{Hri08} of the single transverse proton spin asymmetry $A_{\rm
UT}^{\sin(\phi-\phi_S)}$  versus a tuned (solid) and various ad hoc
GPD model predictions, same as in Fig.~\ref{FigdXSec}.
}
\end{figure}

To provide predictions from our minimalist GPD
models, we rely now on the LO approximation  of
the TFFs (\ref{FFH}), where the effective coupling
is specified in Eq.~(\ref{Cou-Fro}). Thereby, we
employ the Mellin-Barnes integral
(\ref{FFMellinmoments-+}) and the forward Mellin
moments (\ref{GPD-Mod-Min}), which are skewed by
the PW amplitudes (\ref{Mod-H}) and (\ref{Mod-E}).
We like to add that the integration contour must
be chosen carefully and that the dispersion
relations (\ref{GPDs2Re+}) and (\ref{GPDs2Re-})
served us to check our numerics. In the following
we list the various predictions for the
differential cross section (\ref{Pre-CroSec}) that
are displayed together with the HERMES
measurements \cite{Airetal07} in
Fig.~\ref{FigdXSec}, where
statistical and systematical errors are added in
quadrature (same applies for Fig.~\ref{FigTotX}
and Fig.~\ref{FigTPSA}). In the confrontation of
model predictions and measurements we should bear
in mind that at small $-t^\prime \lesssim 0.5\,
\GeV^2$  the longitudinal cross section should
dominate and that the large $-t^\prime$ region, in
which the neglected transverse cross section plays
a role, is a priory not suited for the collinear
factorization approach.\\

\noindent
\underline{dotted:}
{\em pion pole and ${\widetilde H}=0$, running coupling}

\vspace{1mm}

\noindent We recall first the model prediction that relies on an
overwhelming role of the pion pole, shown in Fig.~4 of
Ref.~\cite{Airetal07} as dot-dashed curve. We imitate this model
by setting $\widetilde H$ GPD to zero and replacing $\widetilde {\cal E}$  by
its real part within the parameter set (\ref{Par-Set-Ee1}), see
also left panel in Fig.~\ref{Fig-GPD-2}. The resulting cross section is
plotted in Fig.~\ref{FigdXSec} as dotted line, which resembles the
corresponding curve in Fig.~4 of Ref.~\cite{Airetal07}. As one
realizes the model fails to describe the normalization of the
cross section at very low $- \langle t^\prime \rangle \sim
0.08\,\GeV^2$, where the discrepancy can reach one order of
magnitude at larger $\cQ^2$ values. Since Regge-inspired model
estimates \cite{GuiLagVan97,KasMos09}, e.g., compare dashed and
dotted curves in Fig.~4 of Ref.~\cite{Airetal07}, state that the
$1/R$ ratio should be close to zero at $t^\prime=0$, a popular
conclusion drawn from the observation is that so-called ``power
contributions'' are needed.
\\

\noindent
\underline{dashed:}
{\em pion pole and minimalist $\widetilde{H}$ model,
running coupling}

\vspace{1mm}

\noindent We take now the same model for $\widetilde {\cal E}$ as
above and include the $\widetilde H$  GPD within the parameter set
(\ref{Par-Set-H}), displayed as dashed line. As expected from our
considerations in Sect.~\ref{subsec-GPD-Pro},  the inclusion of
$\widetilde H$ yields a flatter $t$-dependence, compare dashed
with dotted curves. One also realizes that for the lowest $\Bx$
values (left panel), which means also low $\langle \cQ^2 \rangle =
1.4\,\GeV^2$,  the cross section at small $-t^\prime$ increase by
roughly a factor of two and so the discrepancy of model estimate
and measurement diminish, see also the left panel in Fig.~\ref{Fig-GPD-2}.
However, the discrepancy increases with growing $\cQ^2$ and
reaches approximately one order of magnitude at $\langle \cQ^2
\rangle = 5.4\,\GeV^2$. Since the LO prediction of the cross
section is proportional to the square of the running coupling
(\ref{Def-alpha}), its normalization changes within the canonical
scale setting prescription for the renormalization scale $\mu_R=
\cQ$ by
$$
\frac{d\sigma^{\gamma^*p\rightarrow n \pi^+}(\Bx,t,\cQ^2=
5.4\,\GeV^2)}{d\sigma^{\gamma^*p\rightarrow n \pi^+}(\Bx,t,\cQ^2= 1.4\,\GeV^2)}
\sim
\frac{\alpha_s^2(\cQ^2= 5.4\,\GeV^2)}{\alpha_s^2(\cQ^2= 1.4\,\GeV^2)}
\approx 0.5\,.
$$
Hence, we might be tempted to conclude that half of the
discrepancy, observed at $\cQ^2= 5.4\,\GeV^2$,  arises from
logarithmic scaling violation, mainly induced by the running
coupling.\\

\noindent
\underline{dot-dashed:}
{\em pion trajectory $\widetilde{E}$ model
(\ref{Par-Set-Ee1}) and minimalist $\widetilde{H}$ model,
frozen coupling}

\vspace{1mm}

\noindent If we now freeze the coupling (\ref{Cou-Fro}), our model
with reggeized pion pole ($\Im{\rm m} \widetilde\,{\cal E}\neq 0$), describes
data rather well at smaller  $-t^\prime$  values, see
dot-dashed curve. The growing discrepancy within increasing
$-t^\prime$ might be understood as an increase of the transverse
cross section. Indeed, the dot-dashed curve is rather close to the
Regge model prediction \cite{GuiLagVan97} for the longitudinal
cross section, shown in Fig.~4 of Ref.~\cite{Airetal07} as dotted
curve, while the prediction for the full differential cross section, shown
there as dashed curve, matches the experimental measurements.
Hence, if one would borrow the $1/R$ ratio from the Regge-inspired
model, we would presumably conclude that our GPD model describes
data over an astonishing large $|t^\prime|$ range.\\

\noindent
\underline{dot-dot-dashed:} {\em pion trajectory $\widetilde{E}$ model
(\ref{Par-SetE-e}) and minimalist $\widetilde{H}$, frozen coupling}

\vspace{1mm}

\noindent Unfortunately, the parameter set (\ref{Par-Set-Ee1}) is
rather disfavored by lattice data, see dashed curve in the right
panel of Fig.~\ref{figparset-E}. Moreover, as we will see below in
Fig.~\ref{FigTPSA}, it predicts that the single transverse proton spin
asymmetry is of order $-1$ and so it might be considered as ruled
out. Taking the  parameter set (\ref{Par-SetE-e}), which is more
favored by lattice simulations, we obtain the dot-dot-dashed curve
in Fig.~\ref{FigdXSec}. Now the discrepancy of the normalization
at small $t^\prime$ increases again with growing $\cQ^2$. Note
also that the $t^\prime$-dependence is now flatter and if we would
correct the normalization for large value of $\cQ^2$ in the right
panel, the model would perfectly describe data. In other words
our model description would then state that the cross section is
dominated by the longitudinal part of the cross section over a
large lever arm in $-t$. This would then contradict the Regge-inspired
model predictions \cite{GuiLagVan97,KasMos09}.
\\

\noindent
\underline{solid curve:}
{\em pion trajectory $\widetilde{E}$ model (\ref{Par-Set-Ee1}) and
tuning $\widetilde H$, frozen coupling}

\vspace{1mm}

\noindent
As we demonstrated, our minimalist  GPD models provide us with a variety
of predictions and confronting them with experimental
data might imply contradictory conclusions about the failure or
successes of the collinear factorization approach. Hence, as
advocated in Ref.~\cite{KumMuePas08}, it is more appropriate to
change the GPD approach. We now assume that the collinear
factorization approach with a frozen coupling works and ask now
for the functional form of the GPD at its cross-over trajectory,
which entirely determines the imaginary and real part of CFFs to
LO accuracy.  Instead of introducing a new parameterization for
GPDs on the cross-over line we employ here the Mellin-Barnes
integral again. Only for illustration we consider it as
representation for a complex valued function, satisfying the
dispersion relations (\ref{GPDs2Re+}) and (\ref{GPDs2Re-}), where
the physical/partonic meaning of the parameters is mostly lost.
From an "eye fit", where we already included preliminary HERMES
data for the single transverse proton spin,  we find that the
parameters set\\
\begin{eqnarray}
\label{Par-Set-emp}
\widetilde{h}^{\rm val} \stackrel{{\rm ``fit"}}{=}0.98\,,\;\;
\beta^{\rm val}_{\widetilde H} \stackrel{\rm ``fit"}{=} 3/2\,,\;\;
\delta \beta^{\rm val} \stackrel{\rm ``fit"}{=}  1/2\,,
&&
\widetilde{h}^{\rm sea} \stackrel{\rm ``fit"}{=}1\,,\;\;
\beta^{\rm sea}_{\widetilde H} \stackrel{\rm ``fit"}{=} 5\,,\;\;
\delta \beta^{\rm sea} \stackrel{\rm ``fit"}{=}  1\,,\;\;
\\
\widetilde{e}^{\rm val} \stackrel{\rm ``fit"}{=}  0\,,\;\;
\phantom{.98}\beta^{\rm val}_{\widetilde E} \stackrel{\rm ``fit"}{=} 1\,,
\phantom{/2\,,\delta \beta^{\rm val} \stackrel{\rm ``fit"}{=}  1/2\,,}
&&
\widetilde{e}^{\rm sea} \stackrel{\rm ``fit"}{=} 0\,,\;\;
\beta^{\rm sea}_{\widetilde E} \stackrel{\rm ``fit"}{=} 3
\nonumber
\end{eqnarray}
will lead to a good description of data. In Fig.~\ref{FigdXSec} we
realize that the solid curve indeed describes the small
$-t^\prime$ data points and leaves some space for the contribution of the
transverse cross section.

In Fig.~\ref{FigTotX} we also display our model results for the
total cross section $\sigma^{\gamma^*p\rightarrow n \pi^+}$ versus
$\cQ^2$. Such confrontation is sometimes viewed as a test for the
onset of the perturbative approach. However, we should also spell
out some warnings. First at all, the total cross section
$\sigma^{\gamma^*p\rightarrow N M}$ is not a pure twist-two
quantity and secondly even if we would have a measurement of its
longitudinal part $\sigma_L^{\gamma^*p\rightarrow N M}$ it
contains necessarily contributions that arise from phase space
regions for which the collinear factorization approach is perhaps
not applicable. Moreover, since the phase space, accessible in
experiment, is restricted, the kinematical variables might be even
correlated. Naively, one would expect a 1/$\cQ^6$ scaling and if
it would show up in the data one might say that the onset of
scaling justifies the perturbative approach. Also the reversed
logic has been often stressed in the literature. Certainly, in the
first place the running coupling, appearing in the hard amplitude,
and also logarithmic scaling violation, predicted by evolution,
demolishes this logic and reminds us that logarithmic scaling
violations are part of the perturbative framework. Indeed, our two
models within running coupling (dotted and dashed) have a steeper
$\cQ^2$-dependence as the models that were treated with a frozen
coupling (dot-dashed, dot-dot-dashed, and solid) and dashed). The
$\cQ^2$ slope of our ``fitted'' GPD model prediction (solid)
fairly follows the data. That it underestimates the total cross
section measurement is natural, since we describe only the
differential cross section at smaller $-t^\prime$ values. The
differences might arise from the transverse cross section.

We now confront our model predictions in Fig.~\ref{FigTPSA} with
a preliminary measurement of the single transverse proton spin
asymmetry $A_{\rm UT}^{\sin(\phi-\phi_S)} $, which we describe
within the approximation (\ref{Def-AUT}). We realize from this
formula that the asymmetry is proportional to
\begin{eqnarray}
\label{Pro-AUT} A_{\rm UT}^{\sin(\phi-\phi_S)}  \stackrel{{\rm
Tw}-2}{\propto} - \left(\Re {\rm e}\, \widetilde {\cal
E}^{ud}\right) \left(\Im {\rm m}\, \widetilde {\cal H}^{ud }
\right) + \left( \Im {\rm m}\, \widetilde {\cal E}^{ud}\right)
\left(\Re {\rm e}\, \widetilde {\cal H}^{ud }\right)  \,.
\end{eqnarray}
For the pure pion pole model (dotted) the asymmetry trivially
vanishes. In a pion pole model with non-vanishing $\widetilde H$
GPD the asymmetry, shown as dashed curve, is usually predicted to
be negative with a large modulus that is rather flat with respect
to the $t^\prime$-dependence; except that it due to the
kinematical factor $\sqrt{t^\prime/t_{\rm min}}$  in the asymmetry
(\ref{Def-AUT}) steeply vanishes at the kinematical boundary. These
characteristic features arise from the fact that the imaginary
part of the $\widetilde{\cal H}$  CFF, equivalently, the
$\widetilde H$ GPD on the cross-over line, is in all popular GPD
models positive and the imaginary part of $\widetilde {\cal E}$
plays no or not an essential role, see Eq.~(\ref{Pro-AUT}).
Obviously, in a twist-two interpretation such standard GPD models
\cite{FraPobPolStr99,BelMue01a} are ruled out from the preliminary
HERMES measurement \cite{Hri08}.

The dot-dot-dashed curve illustrate that the prediction from the
reggeized pion pole GPD model in the version (\ref{Par-SetE-e})
looses the characteristic $t^\prime$-dependence and it can even
possess a sign change. In this GPD model the real part of the CFF
$\widetilde {\cal E}$ becomes at larger values of $-t^\prime$ in
the vicinity of  $\Bx \sim 0.1$ negative, while $\Im{\rm
m}\widetilde{\cal H}$ remains positive. However, also this
twist-two model prediction is still disfavored by experimental
data.

If we now give up our minimalist model for the GPD
$\widetilde H$ and employ the pragmatic parameterization
(\ref{Par-Set-emp}), the resulting solid curves are in fair
agreement with the experimental points. In this model the
real part of $\widetilde{\cal E}$ is also for larger $-t^\prime$
values positive (in contrast to the dot-dot-dashed model, since we
have now ${\widetilde e}^{\rm val}= {\widetilde e}^{\rm sea} =0$),
while now the imaginary part of $\widetilde{\cal H}$ changes at
$-t^\prime \sim 0.3\,\GeV^2$ the sign and so also the asymmetry.
Note also that both terms on the r.h.s.~of Eq.~(\ref{Pro-AUT}) are
sizeable, where the second term behaves differently than the first
one: it is positive at small $-t^\prime$ values and becomes
negative at larger $-t^\prime$ values. Thus, also delicate
cancellations appear in the asymmetry prediction, which finally
determine the shape of the solid curve.

From the confrontation of our GPD models with the transverse
target spin asymmetry measurements we conclude that a ``predictive
power GPD model approach'', as it has been emphasized for the
overwhelming role of the pion pole, might be misleading. The phase
difference between the `CFF' $\widetilde{\cal H}$  and
$\widetilde{\cal E}$  are not predictable at larger values of
$-t^\prime$. As said before, relying on the twist-two dominance of
this observable one can ``fit'' experimental data to reveal GPDs.
Unfortunately, we have also illustrated that rather different GPD
model scenarios are potentially able to describe the HERMES
measurements.

\begin{figure}[htb]
\centering
\includegraphics[width=17cm]{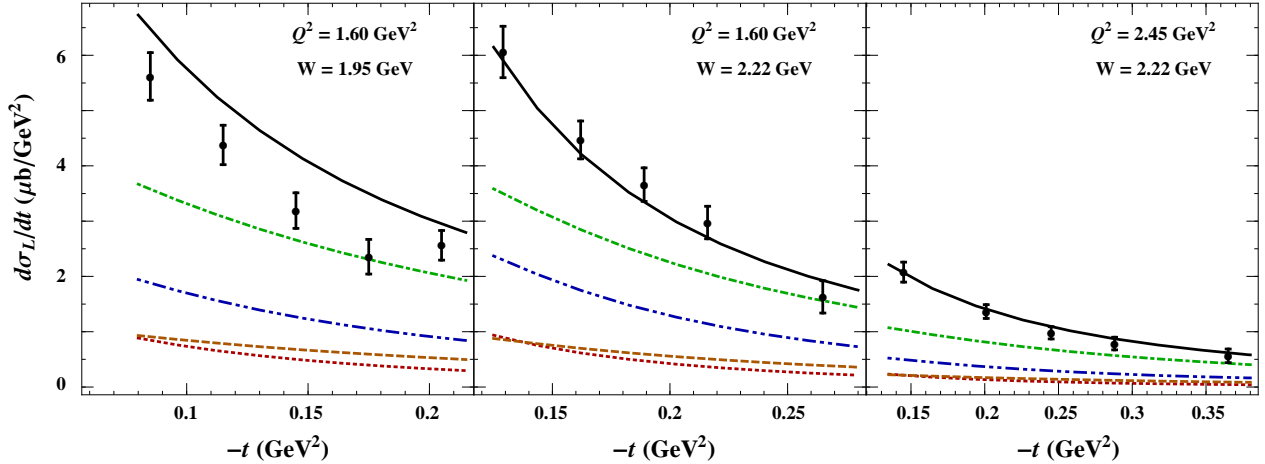}
\caption{\label{FigdXSec-JLAB}\small
Measurements of the longitudinal differential cross section $d\sigma_L^{\gamma^*p\rightarrow n
\pi^+}/dt$ in Hall C at JLAB \cite{Bloetal08} versus various GPD model predictions,
same as in Fig.~\ref{FigdXSec}. Only statistical errors are included.
}
\end{figure}
Finally, we confront  in Fig.~\ref{FigdXSec-JLAB} our models with
the measurements of the longitudinal differential cross section
$d\sigma_L^{\gamma^*p\rightarrow n \pi^+}/dt$ from Hall C
collaboration \cite{Bloetal08}, where we do not account for
systematical errors. As one realizes also this low energy data are
described by the  GPD model (\ref{Par-Set-emp}), ``fitted'' to
HERMES data, while all other models underestimate the measurement.
In particular, the pion pole model (dotted) with running coupling
fails in the normalization and it possesses a too flat
$t$-dependence. Freezing the coupling, the cross section estimate
will be roughly enhanced by a factor five and the pion pole model
without $\widetilde H$ is becoming closer to experimental data.
This estimate is similar to our dash-dotted curve, based on the
$\widetilde E$ parameterization (\ref{Par-Set-Ee1}). However, we
see from the alternative $\widetilde E$ parameterization
(\ref{Par-SetE-e}), shown as dot-dot-dashed curve, that the model
predictions are sensitive to the shape of $\widetilde E$ GPD, too.
Now the model roughly underestimates experimental data by  a
factor three. Employing the parameter set (\ref{Par-Set-emp}), we
enhance the GPDs in the large $x$ region and describe experimental
data. This enhancement feature of cross sections at low energy has
been also observed in the hard $\rho$ electroproduction and it was
modelled within a $D$-like addenda \cite{GuiMor07}. We like to
emphasize the intrinsic duality aspect of GPDs that such an
enhancement is naturally tied to the GPD behavior on the
cross-over line, see also the spectator model \cite{HwaMue07}. We
add that our ``fitted'' GPD model also describes HALL C data from
Ref.~\cite{Horetal07} and that the  VGG plus ``power corrections''
code \cite{VanGuiGui99} is also in fair agreement with HALL C
data, too.

\section{Summary and conclusions}

In this paper we have applied generic GPD modelling
\cite{KumMuePas08a} to set up zero-skewness $\widetilde H$ and
$\widetilde E$ GPDs in the iso-vector sector. Thereby, we were
confronted with the problem that Regge phenomenology within
unnatural parity exchanges is challenging. Relying on the
restoration of chiral symmetry in hard or even high-energy
processes, we conjectured a master trajectory for odd $J^{++}$
exchanges. Such a restoration might be also somehow indicated by
the appearance of the $a_3(1875)$  meson, which is listed by the
PDG under further states. The hypothetic trajectory naturally
explains the rise of the polarized DIS structure function $g_1$
at smaller $\Bx$.  We illustratively took a SU(6) symmetric
valence scenario for the net contribution.  In such a scenario the
`spin puzzle' in the iso-vector valence quark sector is trivially
resolved. The incorporation of $t$-dependence in our generic GPD
model is based on a Regge pole product ansatz in Mellin space,
where the PCAC Ward-identity could be satisfied within a linear
pion trajectory. Both form factor and generalized form factor
lattice measurements are then well described. Certainly, the model
can be refined, were in accordance with spectroscopic
data a non-linear pion trajectory should be taken into account.

Generic arguments to skew a GPD and to find so its value on the
cross-over line are only given by large-$x$ counting rules. Compared to
the skewness-zero GPD, they are enhanced in the large-$x$ region.
Moreover, the skewness dependence of GPDs could be also
constrained by precise lattice measurements, expected to be
available somewhen in future, however, it is clear that this
information is not sufficient to pin down a GPD on the cross-over
line. Here we set up ad hoc models within the minimalist skewing
prescription in which only the leading SO(3) partial wave in the
expansion of conformal moments is taken into account. Our GPD
models on the cross-over line possess the expected large-$x$
behavior, however, the suppression of $t$-dependence for $x\to 1$
might be too weak. More flexible models can be built by including
non-leading SO(3) partial waves and they can be tuned to
experimental data  \cite{KumMue09}.

To confront the ad hoc GPD models with experimental measurements,
we relied on the collinear factorization approach to LO accuracy.
Thereby, perturbative corrections were taken qualitatively into
account by incorporating them in the coupling via the BLM scale
setting prescription. This might to some extent cure  the
perturbative approach. However, the coupling itself is now
considered as a {\em non-perturbative} expansion parameter that must be
fixed from experimental data, too. In such a procedure it is
rather likely that the coupling  looses its universality  and it
should be rather considered as an effective one. Nevertheless, we
end up with a semi-quantitative framework for the GPD
phenomenology that can be tested versus experimental measurements.

We demonstrated then that the hard $\pi^+$ production cross
section for HERMES kinematics at small $-t^{\prime}$ can be
described within our minimalist GPD models. Thereby, the
$\widetilde H$ GPD is important, too, and so the pion pole in the
$\widetilde E$ GPD plays an essential, however, not an
overwhelming role.  We also considered the single transverse
proton spin asymmetry, preliminarily measured by the HERMES
collaboration, where we assumed that power suppressed
contaminations, induced by the exchange of transverse polarized
photons, are small. For small $-t^\prime$ values the sign of the
asymmetry is determined by both the pion pole and the sign of the
$\widetilde H$ GPD on the cross-over line. In our ad hoc models
the sign of both contributions is positive, in agreement with the
experimental findings. However, or ad-hoc models are now
disfavored for larger $-t^\prime$ values, where they predict as
other popular models a large negative asymmetry. Tuning the GPDs
on the cross-over line, we were able to describe the measurements, too.
For larger $-t^\prime$  the pion pole is not overwhelming and the
relative phase between $\widetilde{\cal H}$ and $\widetilde{\cal
E}$ `CFFs' has an intricate variable dependence, which can hardly
be predicted. In our perturbative framework the same ad hoc models
are also disfavored by the longitudinal photoproduction cross
section measurements in HALL C at JLAB.

Our phenomenological studies illuminated the problems in the
extraction of GPDs from hard exclusive meson electroproduction
measurements. Finally, let us clearly spell out the challenges. On
one hand GPDs are intricate functions, depending on the momentum
fraction, the skewness parameter, and the momentum transfer
squared. Thus, one can not expect that popular GPD models, which
we mainly consider to be ad hoc, are able to describe experimental
data. This problem%
\footnote{Alternatively, one might first ask for the value of CFFs
by utilizing analytic formulae \cite{BelMueKir01,BelMue08} or
fitter codes \cite{Gui08a,GuiMou09}, where the later method is
accomplished with error estimates.} can be resolved within
flexible GPD models used in robust numerical fitting routines
\cite{KumMuePas07,KumMue09}. So far this approach has been applied
to deeply virtual Compton scattering in the small-$\Bx$ region and
within `dispersion relation' technique for fixed target kinematics
\cite{KumMue09}. Indeed, such an approach allowed for the first
time to access the $H$ GPD at leading order and revealed so the
failure of ad hoc models, used before. We emphasize that for the
hard exclusive meson electroproduction the theoretical framework
is not well understood. To judge on the collinear factorization
approach and its improvements, one can utilize the universality of
GPDs. Namely, from a widely phenomenological application of the
GPD approach within flexible GPD models or `dispersion relations'
one might test its internal consistency. If such a
phenomenological consistency check is successful, one
simultaneously reveals the GPDs, mainly on the cross-over line.

\section*{Acknowledgements}

D.M.~is grateful to K.~Kumeri{\v c}ki and K.~Passek-Kumeri{\v c}ki
for the collaboration in generic modelling of parity-even GPDs,
where the framework has been adopted here to the parity-odd sector. We are indebted to
I.~Hristova for  providing us preliminary HERMES data and to
J.~M.~Laget for numerical model results. For discussions on Regge
and model aspects we thank S.~S.~Afonin, M.~Polyakov, M.~Guidal,
and P.~Kroll.

{\ }

\noindent
{\bf First note added}

\noindent
During the preparation of the manuscript we were becoming
aware of (preliminary) photon-to-pion transition form factor
measurements from the BaBar collaboration, which are now
available in Ref.~\cite{Aubetal09}. These experimental data do not support
the canonical scaling, which we have implicitly assumed, to
fix the inverse moment of the pion distribution amplitude.
They challenge our interpretation of the connection among
photon-to-pion form factor, pion form factor, and hard
exclusive $\pi^+$ electroproduction, however, the
normalization of our numerical results is essentially
adopted from the pion form factor. Nevertheless, the BaBar
measurement \cite{Aubetal09} yield a `crisis' for the perturbative
description of {\rm pion related} processes and the present
understanding of the partonic substructure of the {\rm
pion}. First comments and interpretations of the  BaBar measurement
are given in Refs.~\cite{MikSte09,Rad09,Pol09}. Since the
pion-to-photon transition form factor is a key process for a
QCD understanding, we would like to encourage the BELLE
collaboration to cross check the measurement and we recall
that a double tagged experiment might provide a deeper
insight into the `pion crisis'. We add that the theoretical
description of all exclusive B decay into pion channels  is
affected by this `crisis'.   \\

\noindent
{\bf Second note added}

\noindent
After our  manuscript was essentially finished, S.V.~Goloskokov
and P.~Kroll submitted their work {\em An attempt to understand
exclusive $\pi^+$ electroproduction} \cite{GolKro09}. The authors
rely on a `hand-bag' approach, which can be considered as a model
improved collinear GPD approach. The advantage of such an approach
is that various observables can be successfully described, which
provides some trust in the underlying GPD framework. We try to
stay as close as possible to the collinear factorization approach
in the hope that  GPD universality and a systematical improvement
can be ensured in a maximal manner. This we consider as a
requirement that allows aiming for the primary goal, namely, to
reveal GPDs from experimental measurements.


\end{document}